\documentclass[fleqn,twoside]{article}
\usepackage{epsfig}
\usepackage[headings]{myespcrc2}

\readRCS
$Id: espcrc2.tex,v 1.2 2004/02/24 11:22:11 spepping Exp $
\usepackage{graphicx}

\newcommand {\lapprox}
   {\, \raisebox{-0.7ex}{$\stackrel {\textstyle<}{\sim}$} \,}

\newcommand{\AmS}{{\protect\the\textfont2
  A\kern-.1667em\lower.5ex\hbox{M}\kern-.125emS}}

\title{Deep Inelastic Scattering at the TeV Energy Scale and the LHeC 
Project}

\author{Paul Newman\address[MCSD]{School of Physics \& Astronomy, 
University of Birmingham, B15 2TT, UK.}}
       
\runtitle{DIS at the TeV scale and the LHeC}
\runauthor{Paul Newman}

\begin{document}

\begin{abstract}
The prospect of an $ep$ collider involving an
LHC proton beam and a new electron accelerator is discussed.
Configurations reaching centre of mass energies a factor
of 5 beyond HERA are possible with luminosities of the
order of $10^{33} \ {\rm cm^{-2} s^{-1}}$. 
The physics programme with such a facility is surveyed
and possible machine and detector lay-outs are sketched. 
\vspace{1pc}
\end{abstract}

\maketitle

\section{INTRODUCTION}

As is clear from the varied contributions to this
workshop concerned with ongoing work on HERA data, 
much is still being learned from the world's 
first electron proton ($ep$) collider \cite{reviews}.
Measurements based on the final HERA data remain of high importance
for the physics of strong interactions and proton structure
in general and
in particular for applications at the 
LHC \cite{HERA:LHC}. By now, our understanding of many 
small $Q^2$ and low $x$ phenomena in Deep Inelastic
Scattering (DIS) is
limited by experimental systematics and 
theoretical uncertainties, rather than by statistical precision. 
However, particularly for searches for
new particles 
and for the understanding of electroweak
effects, the principle limitation is often the available integrated
luminosity.

Without doubt, 
energy-frontier physics will be dominated for
the foreseeable future by the proton and heavy ion beams of the LHC, 
whose unprecedented energy and intensity 
herald a new era in the field. It is 
reasonable to ask whether
these proton beams could be exploited as part of a new high 
performance $ep$ and electron-ion ($eA$)
`Large Hadron electron Collider' (LHeC) \cite{ferdi,LHeC:web}, 
complementing the LHC $pp$ and 
$AA$ programme and a possible pure lepton collider at the TeV energy 
scale. 

\begin{figure}[h] \unitlength 1mm
  \begin{center}
    \begin{picture}(50,57)
      \put(-12,-10){\epsfig{file=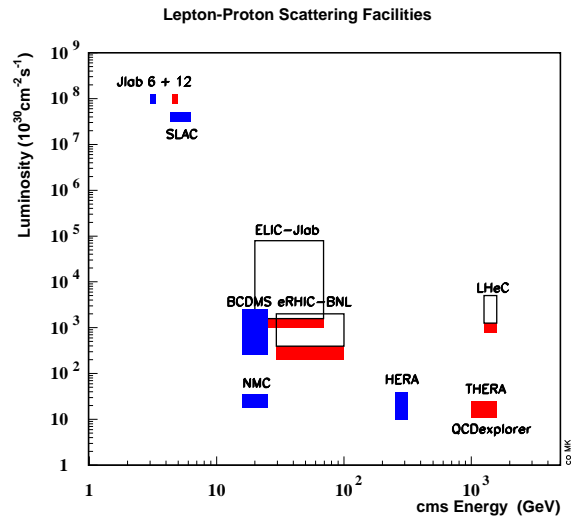,height=0.425\textwidth}}
    \end{picture}
  \end{center}
  \caption[]{Plot of energy versus luminosity for various $ep$ facilities
which have already been realised (blue) and which have 
not (red) \cite{eic,previous:ep}. Open
areas correspond to possible upgrade scenarios. The values shown for
the LHeC are for the configuration described in \cite{ferdi}.}
\label{energy:lumi}
\end{figure}

\begin{figure*}[tb] \unitlength 1mm
  \begin{center}
    \begin{picture}(100,52)
      \put(-25,-12){\epsfig{file=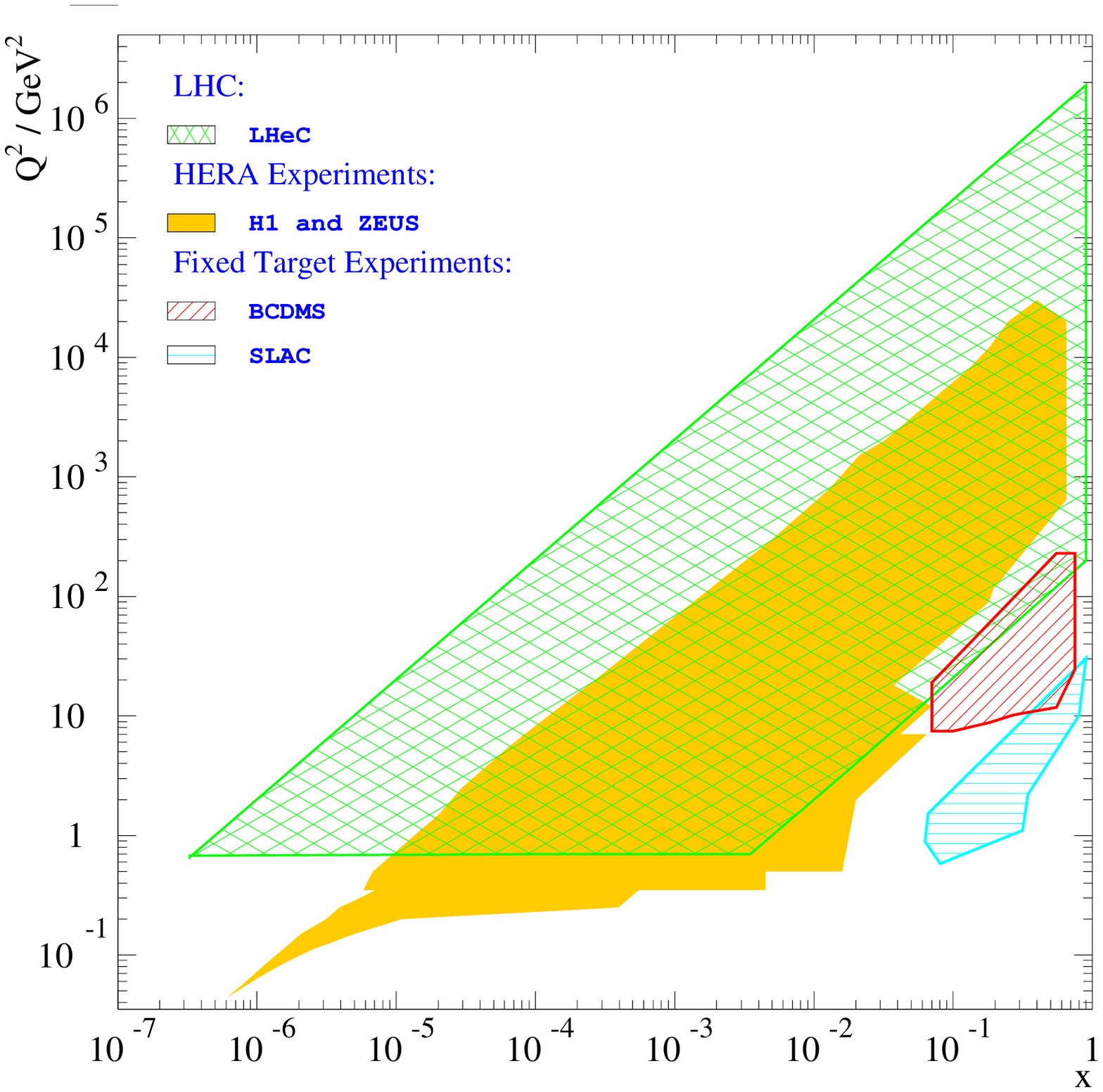,height=0.4054\textwidth}}
      \put(53,-17){\epsfig{file=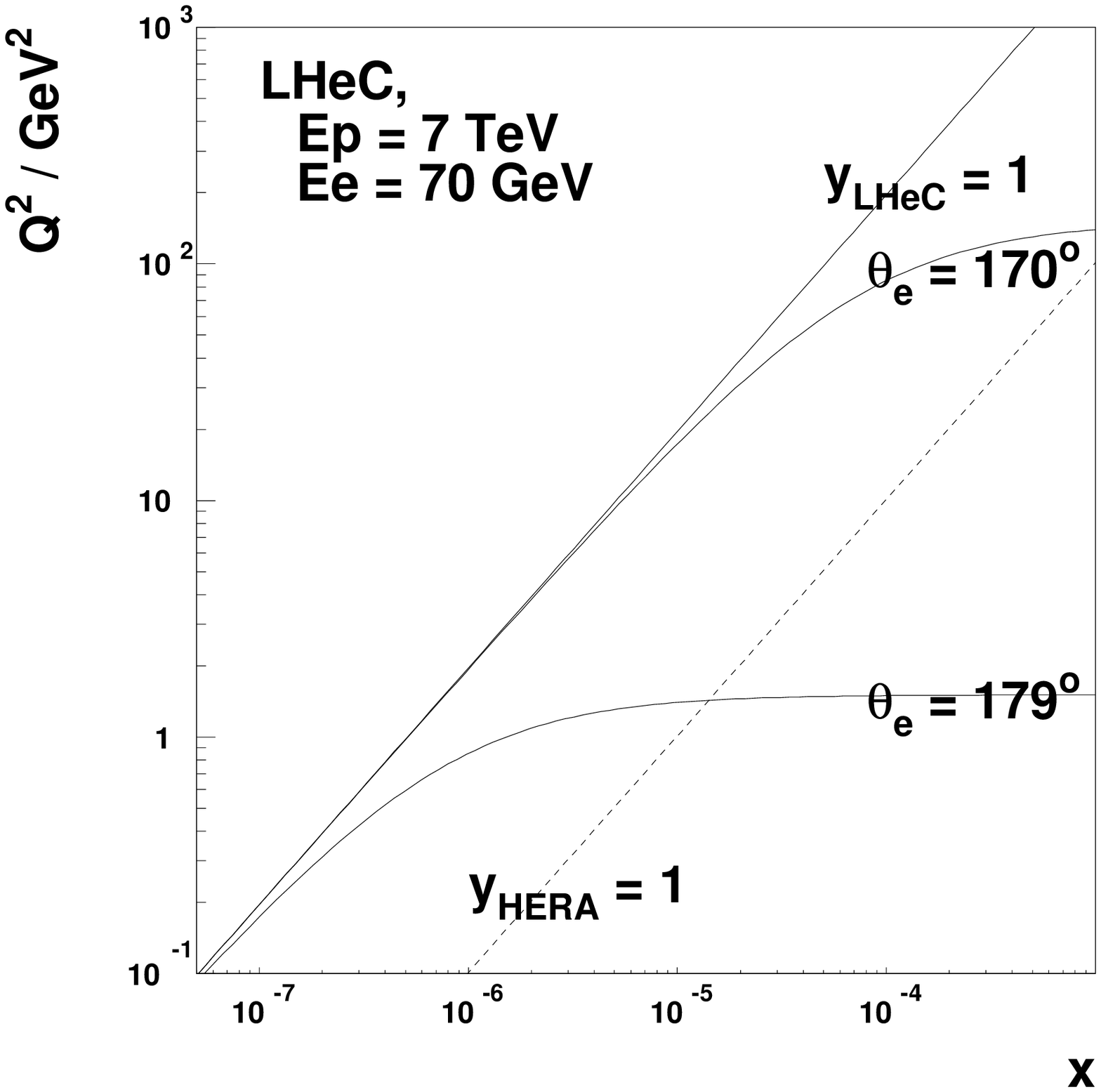,height=0.4744\textwidth}}
      \put(-12,22){\Large{\bf{(a)}}}
      \put(72,22){\Large{\bf{(b)}}}
    \end{picture}
  \end{center}
  \caption[]{(a) Kinematic plane for $ep$ collisions,
showing the coverage of fixed target
experiments, HERA and an LHeC with $70 \ {\rm GeV}$ electrons.
(b) Zoomed view of the low $x$ corner of the kinematic
plane, showing the $y = 1$ kinematic limits at HERA and the LHeC
and the acceptances for two different cuts on electron scattering
angle $\theta_e$ at the LHeC.}
\label{kinplane}
\end{figure*}

A 2006 study of the possibility of adding an electron 
beam to the LHC complex \cite{ferdi} 
suggested that $ep$ collisions with an electron  
energy $E_e = 70 \ {\rm GeV}$ 
and a luminosity of order $10^{33} \ {\rm cm^{-2} s^{-1}}$ 
could be achieved
at moderate power consumption. This would yield a
centre of mass energy of 
$1.4 \ {\rm TeV}$ and would probe distance scales of the order of 
$10^{-20} \ {\rm m}$. For comparison, the best performance achieved 
at HERA was a luminosity of $5 \times 10^{31} \ {\rm cm^{-2} s^{-1}}$ 
at an $ep$ centre of mass energy of $318 \ {\rm GeV}$. 
The large luminosity increase in particular sets the LHeC aside from 
previously evaluated possible future 
high energy $ep$ colliders \cite{previous:ep} 
(figure~\ref{energy:lumi}). If realised, 
an $ep$ machine with this performance 
would lead to the first precise study of ${\rm TeV}$-scale lepton-quark 
interactions.

On September 1-3 2008, a diverse mixture of approximately 90
accelerator scientists, experimentalists and theorists 
met at Esplanade du Lac, Divonne, near CERN, for the inaugural 
meeting \cite{divonne} 
of the ECFA-CERN commissioned LHeC workshop. 
The aim is to assess the physics potential of an electron 
beam interacting with LHC protons and ions as well as its accelerator, 
interaction region and detector requirements and the impact on the existing 
LHC programme.
This contribution is intended both as a summary of
the Divonne meeting and as a snapshot of the current status of the 
LHeC project, with the 
focus mainly on the physics motivation.


\section{KINEMATICS AND GEOMETRY}
\label{geometry}

The accessible kinematic plane assuming a $7 \ {\rm TeV}$ proton and a
$70 \ {\rm GeV}$ electron beam is 
compared with previous experiments in figure~\ref{kinplane}a.
The coverage is extended compared with HERA 
towards low Bjorken $x$ at fixed $Q^2$ or towards high $Q^2$ at fixed $x$
by the ratio of
squared  centre of mass energies  
$s_{_{\rm LHeC}} / s_{_{\rm HERA}} \sim 20$.
With sufficient luminosity to overcome the basic $1/Q^4$ cross
section dependence, squared 4-momentum transfers 
$Q^2 \sim 10^6 \ {\rm GeV^2}$ are accessible. 
As well as sensitivity to new physics (section~\ref{exotics}), 
the high luminosity will
clarify many issues with parton distribution functions (PDFs)
particularly at the highest $x$ 
(section~\ref{partons}). The
ultra-high parton density
region $x \lapprox 10^{-4}$ will be accessed for the first time
at sufficiently large $Q^2$ for perturbative QCD techniques to be
applied 
(sections~\ref{lowx},~\ref{diffraction} and~\ref{ions}). 
When the LHC runs with heavy ions, the
LHeC becomes the first ever $eA$ colliding beam machine, extending
the known kinematic plane for nuclear structure functions by
four orders of magnitude
(section~\ref{ions}).

\begin{figure*}[tb] \unitlength 1mm
  \begin{center}
    \begin{picture}(100,47)
      \put(-30,28){\epsfig{file=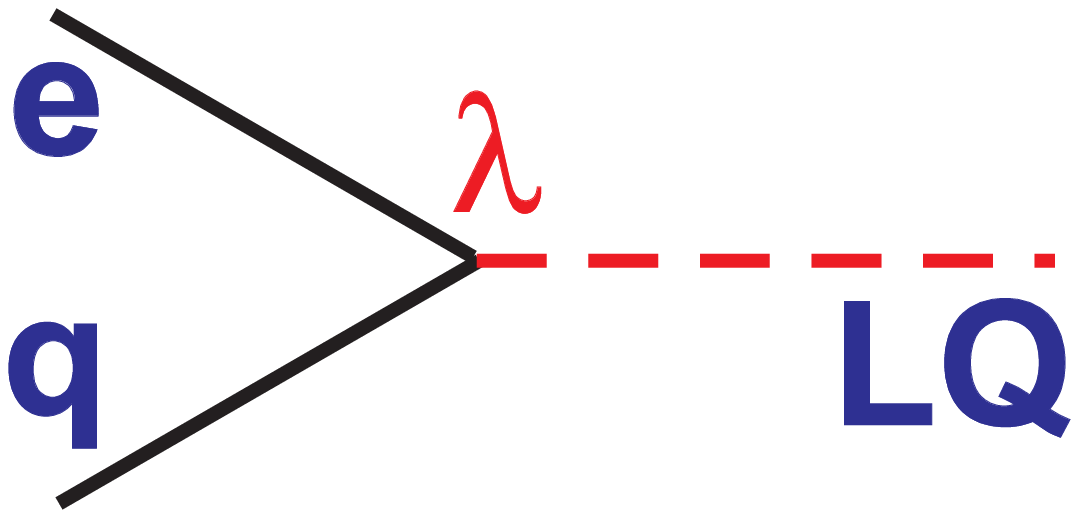,height=0.08\textwidth}}
      \put(-30,-6){\epsfig{file=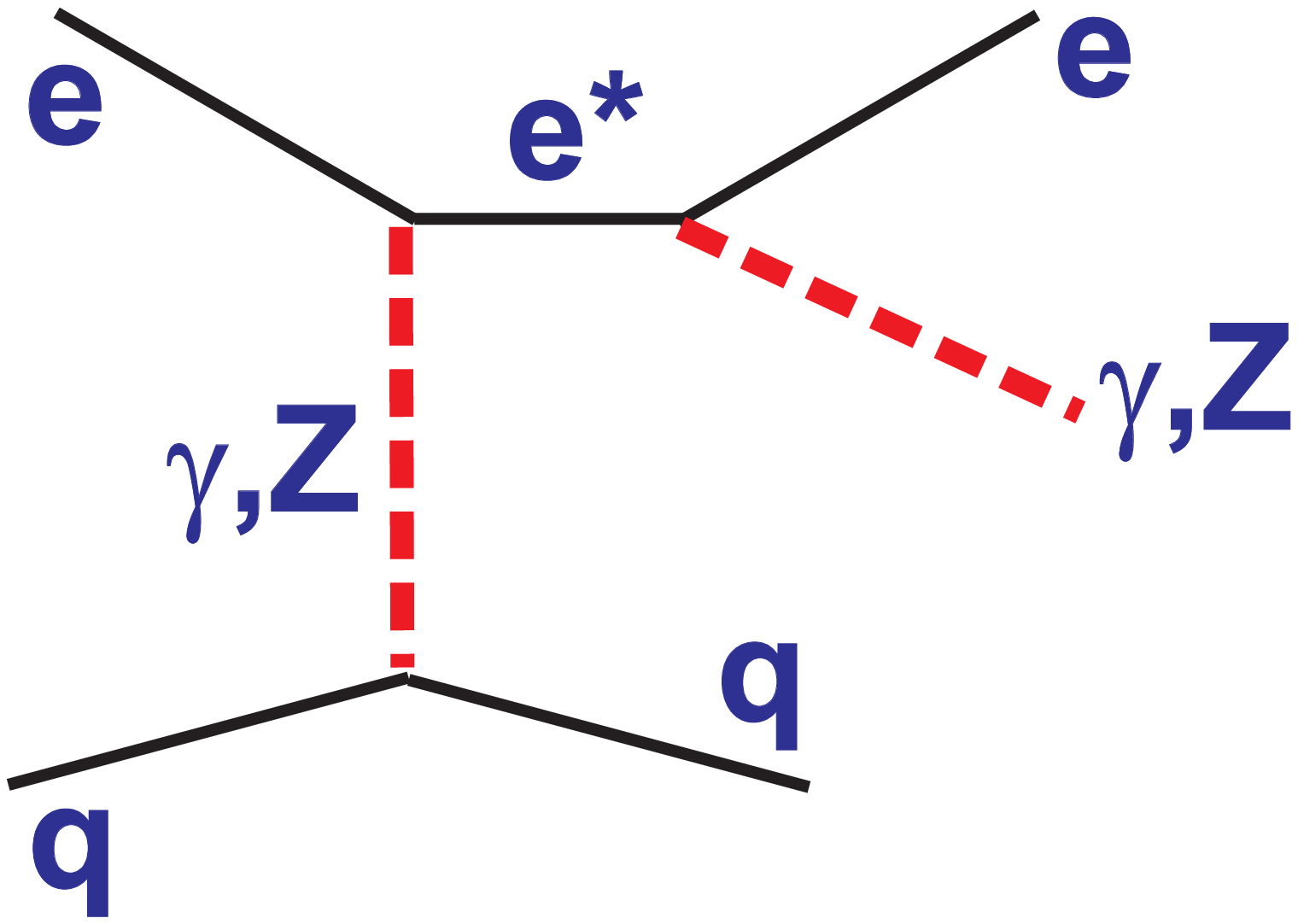,height=0.13\textwidth}}
      \put(0,-10){\epsfig{file=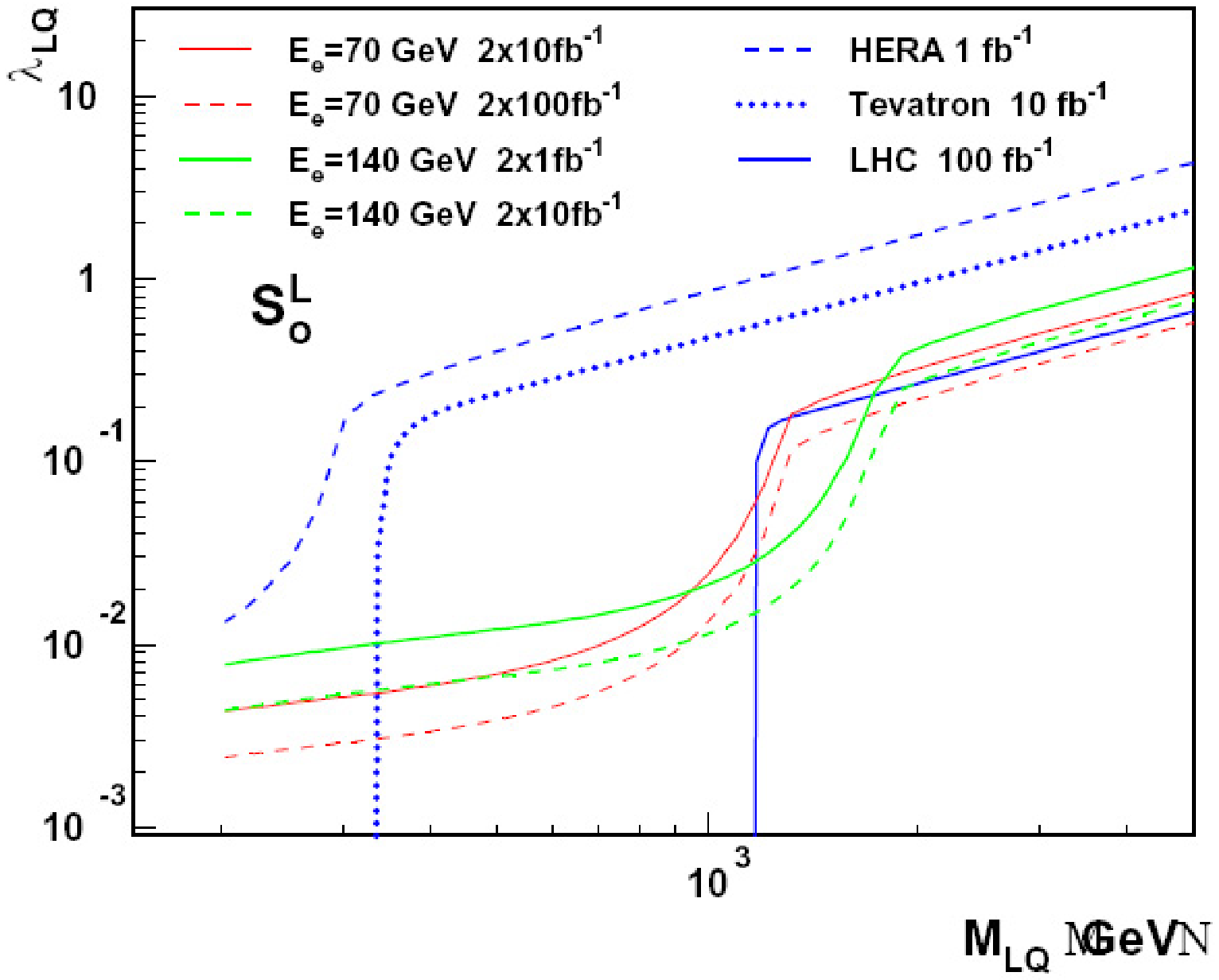,height=0.35\textwidth}}
      \put(68,-8){\epsfig{file=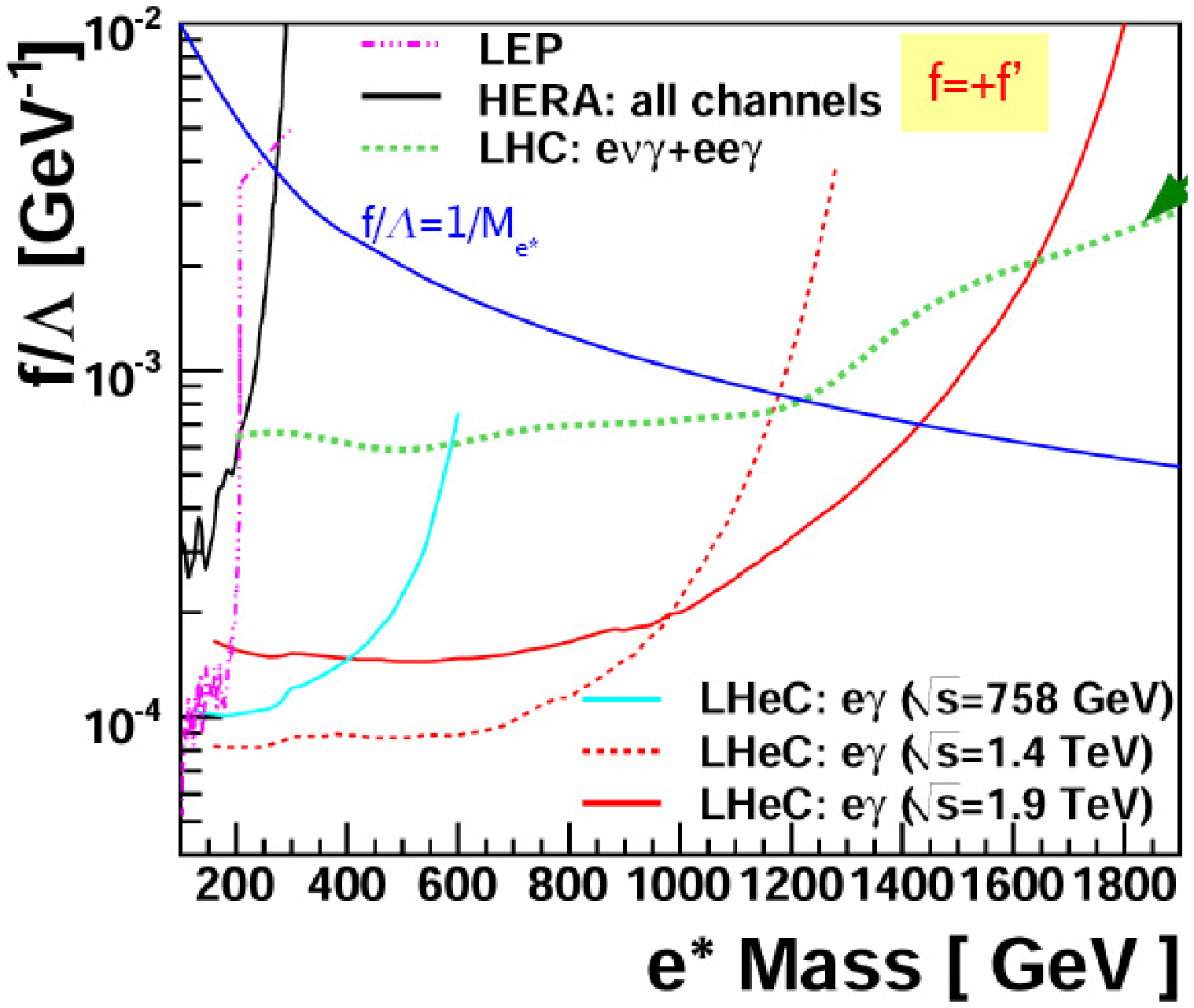,height=0.33\textwidth}}
      \put(-26,43){\Large{\bf{(a)}}}
      \put(12,20){\Large{\bf{(b)}}}
      \put(-26,17){\Large{\bf{(c)}}}
      \put(83,20){\Large{\bf{(d)}}}
    \end{picture}
  \end{center}
  \caption[]{(a) Illustration of the single leptoquark $ep$ production vertex
via a Yukawa coupling $\lambda$. (b) Example comparison between expected
leptoquark limits at the LHC ($200 \ {\rm fb^{-1}}$)
and the LHeC ($2 - 200 \ {\rm fb^{-1}}$ as indicated) \cite{zarnecki}. 
(c) Illustration
of excited electron production in $ep$ interactions. (d) Comparison between 
expected excited electron limits at the LHC ($100 \ {\rm fb^{-1}}$)
and in $e \gamma$ final
states the LHeC ($10 \ {\rm fb^{-1}}$ at $E_e = 20 \ {\rm GeV}$ and 
$70 \ {\rm GeV}$ or $1 \ {\rm fb^{-1}}$ at $E_e = 140 \ {\rm GeV}$) 
\cite{trinh}.}
\label{exotic:fig}
\end{figure*}

Accessing the full available phase space brings challenges in the
detector and interaction region design, as illustrated for the 
example of the scattered electron kinematics in figure~\ref{kinplane}b. If the 
electron detection acceptance extends to scatterings through a 
$1^\circ$ angle ($\theta_e = 179^\circ$)\footnote{The coordinate
system assumed here follows that from HERA, with the $+z$ axis
and `forward' direction corresponding to that of the outgoing
proton beam. Polar angles $\theta$ are measured with respect to
this direction.}, full coverage of the region $Q^2 > 1 \ {\rm GeV^2}$
is obtained, reaching below $x = 10^{-6}$. In contrast, 
with detector components restricted to $\theta < 170^\circ$, 
there is little acceptance for $Q^2 < 100 \ {\rm GeV^2}$ or
$x < 10^{-4}$. Optimising the luminosity by including beam
focusing elements close to the interaction region \cite{ferdi,divonne}, 
similar to those installed for the upgrade from HERA-I to HERA-II,
must therefore be evaluated against the 
implications of the corresponding loss of
small angle detector acceptance.

In the physics studies presented in section~\ref{physics}, two basic scenarios
are considered,
corresponding to 
rough estimates at performance (see section~\ref{experiments} 
for more
details). In the first, beam focusing magnets fill the 
region $\theta > 170^{\circ}$, allowing an integrated luminosity of nominally
$10 \ {\rm fb^{-1}}$ per year to be achieved. In the second, it
is assumed that there are no focusing magnets, allowing acceptance
to $179^{\circ}$, but reducing the annual luminosity by an order
of magnitude to $1 \ {\rm fb^{-1}}$. In both cases, 
unless otherwise stated, a $70 \ {\rm GeV}$ electron beam is
assumed. In estimating systematic precision, modest, 
factor-of-two, improvements
over the performance of the HERA detectors are assumed,
with electromagnetic and hadronic energy scale uncertainties
of $0.1 \%$ (at the kinematic peak) and $0.5\%$,
respectively, and a polar angle alignment good to 
$0.1 \ {\rm mrad}$ \cite{max:dis07}.

\section{PHYSICS POTENTIAL}
\label{physics}

\subsection{Rare and exotic Processes}
\label{exotics}

As discussed for many years \cite{salam,dainton}, 
it is natural to
think - at some fundamental level - of quarks
and leptons as different low energy manifestations of a 
single underlying form of matter, explaining the 
quark-lepton symmetry of the Standard Model, but contrasting
with the clear distinctions between them at currently accessible scales.
When searching for new physics, the
electron-quark vertex therefore deserves precision study
in parallel with pure-lepton and pure-strongly-interacting vertices. 
Whilst the LHeC does not have anything like the discovery potential 
of the LHC, it gives interesting sensitivity to new 
leptons or particles with 
both lepton and baryon quantum numbers. It also complements the LHC search
potential by offering a relatively clean environment in
which to understand the nature and details of its new discoveries.

\begin{figure*}[tb] \unitlength 1mm
  \begin{center}
    \begin{picture}(100,33)
      \put(-25,-5){\epsfig{file=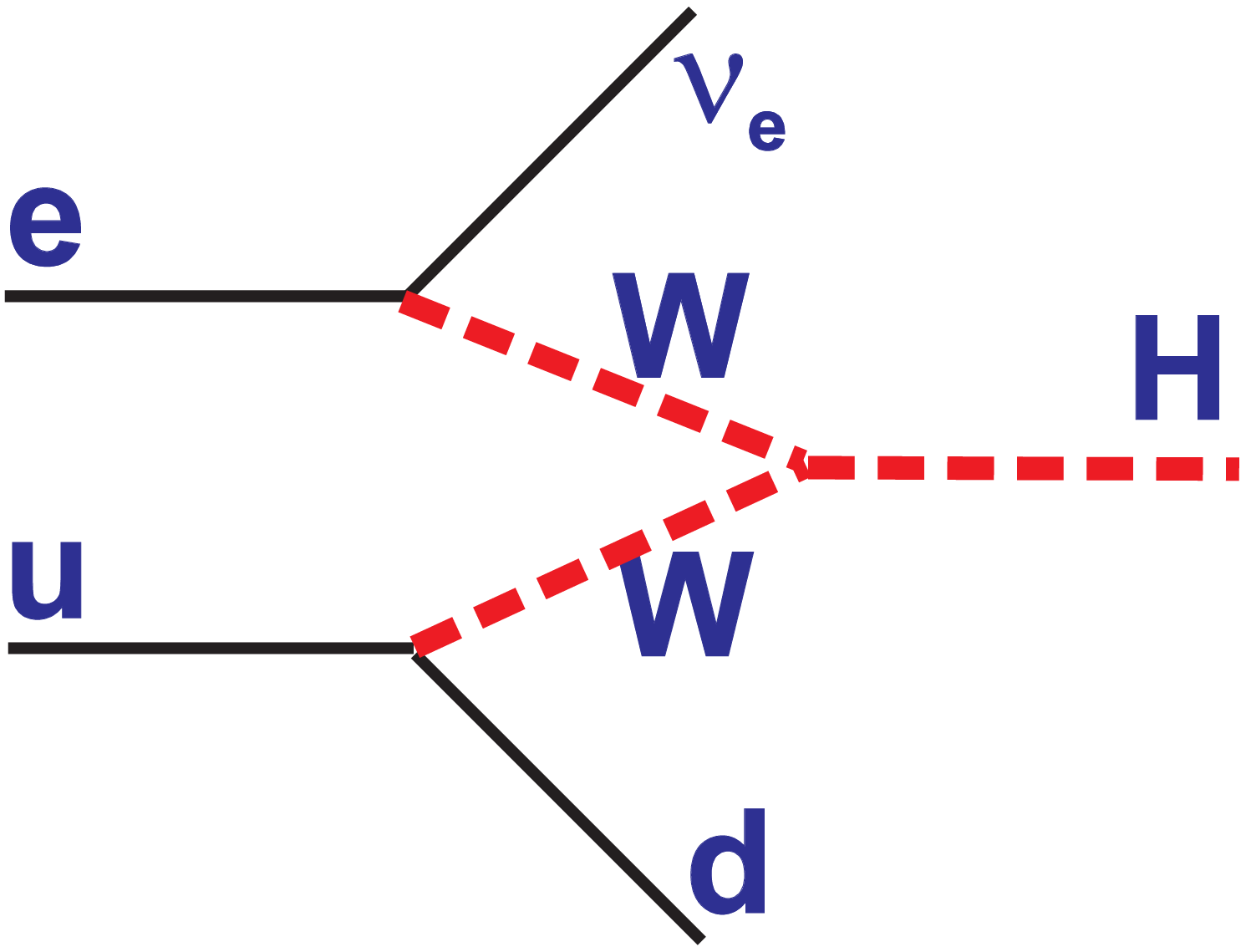,height=0.2\textwidth}}
      \put(30,-13){\epsfig{file=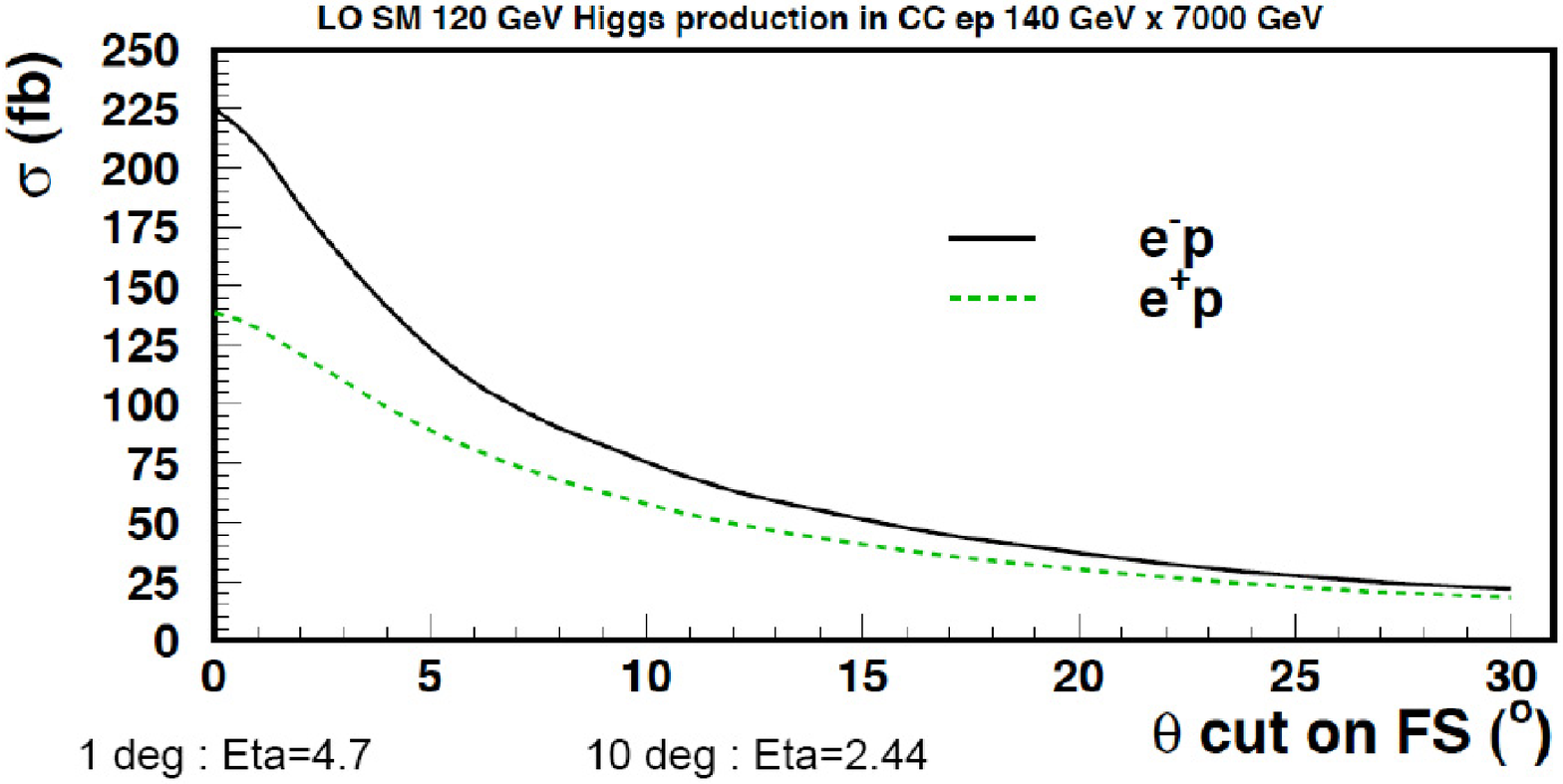,height=0.3\textwidth}}
      \put(-26,25){\Large{\bf{(a)}}}
      \put(110,25){\Large{\bf{(b)}}}
    \end{picture}
  \end{center}
  \caption[]{(a) Illustration of the dominant LHeC Higgs production 
mechanism. (b) Integrated `visible' Higgs production cross sections
for $e^+p$ and $e^-p$ scattering 
as a function of the forward limit of the detector acceptance for an
example case of $E_e = 140 \ {\rm GeV}$ \cite{uta}.}
\label{higgs}
\end{figure*}

An LHeC 
could be uniquely sensitive to the physics of
massive new electron-quark 
bound states (leptoquarks, figure~\ref{exotic:fig}a), 
which exist in a variety of models,
such as $R$-parity violating supersymmetry. 
As illustrated for the example of a scalar
leptoquark of zero fermion number in figure~\ref{exotic:fig}b \cite{zarnecki},
the leptoquark mass range covered by a search in $20 \ {\rm fb^{-1}}$ of LHeC
data is comparable to that accessed with $100 \ {\rm fb^{-1}}$ at the LHC.
However, since leptoquarks are almost always pair-produced at the 
LHC, unravelling their potentially complex spectroscopy \cite{buchmueller}
would be difficult. 
The more easily controlled single production vertex in $ep$ physics  
allows fermion number to be determined for example from 
electron beam charge 
asymmetries, spins from angular decay distributions and 
chirality from beam polarisation asymmetries \cite{ferdi}. 

Another exotic scenario to which the initial state lepton at the 
LHeC gives high sensitivity is the production of excited leptons,
as illustrated in 
figure~\ref{exotic:fig}c. The expected
limits from studying $e \gamma$ final states in
one year of LHeC data are shown in 
figure~\ref{exotic:fig}d \cite{trinh}. For a wide range of excited electron
masses, the LHeC sensitivity extends to significantly lower
$e^* \rightarrow e \gamma$ couplings than are accessed at the 
LHC. Similarly high sensitivity to 
excited neutrinos is available from $\nu \gamma$ final states.

Suspersymmetric electrons are another area of high
sensitivity. The process $eq \rightarrow \tilde{e} \tilde{q}$
via neutralino exchange is within observable LHeC range for sums
of the selectron and squark masses up to around $1 \ {\rm TeV}$. 
If relatively light squarks are observed at the LHC and the 
selectron is heavy, the LHeC sensitivity would be competitive 
with that of the LHC \cite{selectron}.

Beyond searches for new particles, 
the LHeC would complement the LHC in the investigation of the 
electroweak sector of the Standard Model. 
Top quarks would be produced copiously, both singly and in pairs, 
in the relatively clean environment offered by $ep$ scattering \cite{top}.
Light Standard Model Higgs bosons would be 
produced dominantly through $WW$ fusion (figure~\ref{higgs}a)
with potentially around 1000 events per
year \cite{uta}. A light Higgs could be studied in the
dominant $b \bar{b}$ decay
mode, which is problematic at the 
LHC. 
The visible cross section for 
$\theta > 1^{\circ}$ is around twice that for
$\theta > 10^{\circ}$ (figure~\ref{higgs}b).

\subsection{Parton densities}
\label{partons}

With LHeC data, the proton PDFs could be 
measured at previously unexplored $Q^2$ values beyond 
$10^6 \ {\rm GeV^2}$ and at small
$x$ values, approaching $10^{-6}$ in the DIS region. The $x$ range covers
that required for a full 
understanding of the initial state of
parton-parton scattering on the rapidity 
plateau at the LHC. 

\begin{figure*}[tb] \unitlength 1mm
  \begin{center}
    \begin{picture}(100,43)
      \put(-20,-13){\epsfig{file=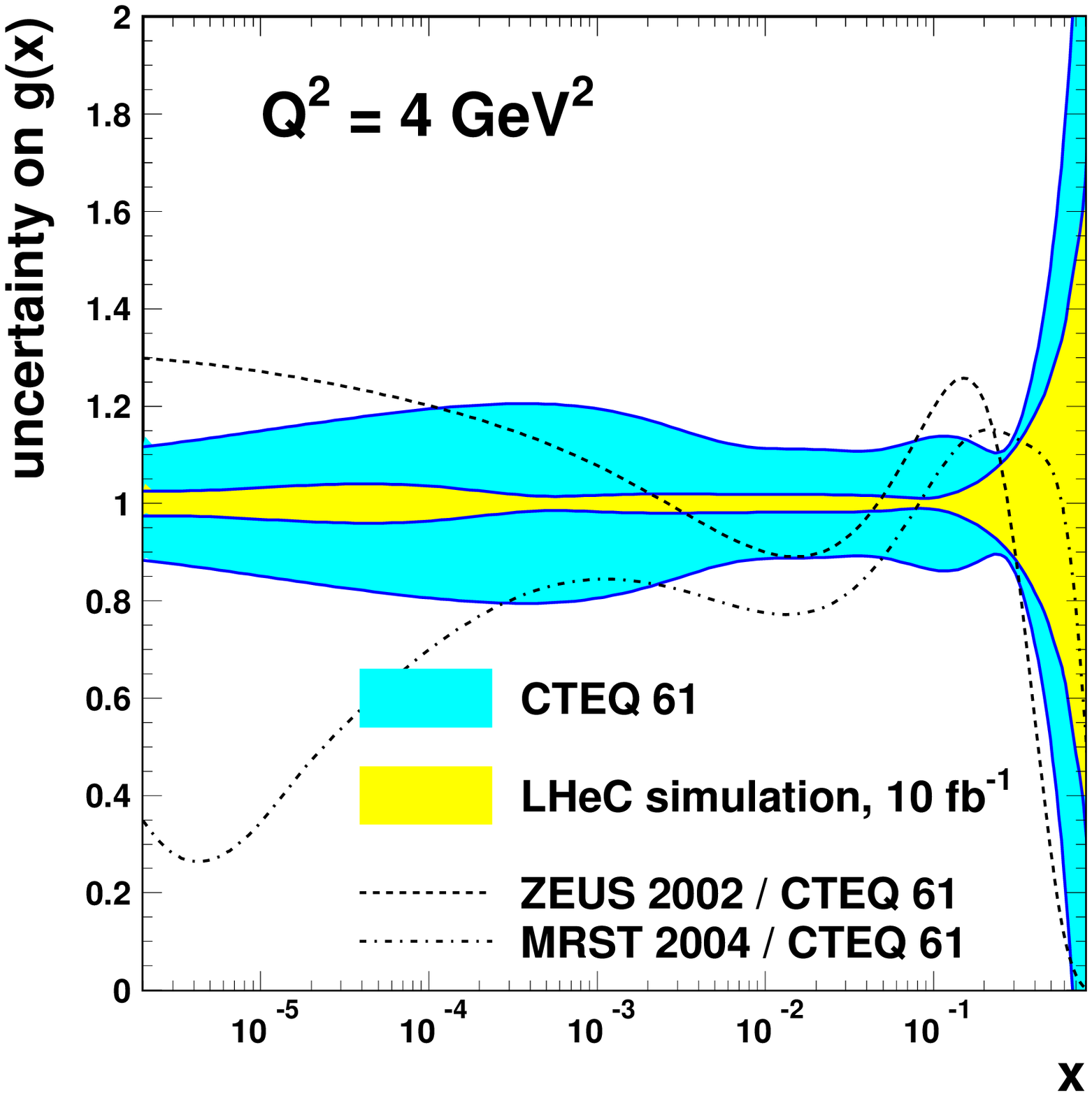,height=0.42\textwidth}}
      \put(60,-13){\epsfig{file=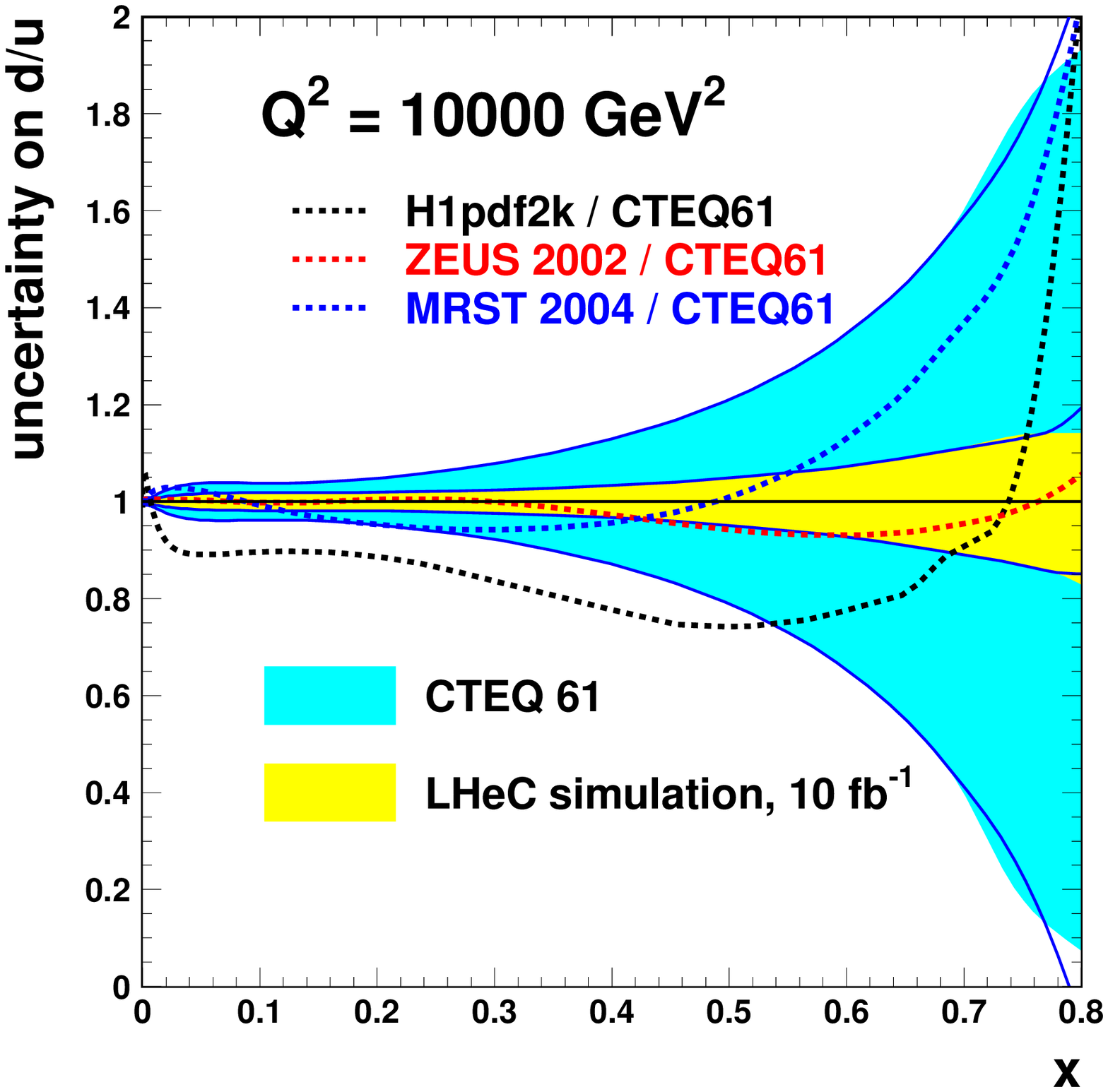,height=0.42\textwidth}}
      \put(24,38){\Large{\bf{(a)}}}
      \put(106,38){\Large{\bf{(b)}}}
    \end{picture}
  \end{center}
  \caption[]{Comparisons 
with current results from global
fits by the CTeQ group \cite{cteq6}
of predicted LHeC precision on (a) the gluon
density at $Q^2 = 4 \ {\rm GeV^2}$ and (b) the $d/u$ ratio at
$Q^2 = 10\,000 \ {\rm GeV^2}$ \cite{max:dis07}. The differences relative to 
other recent parton density parameterisations are also shown.}
\label{pdf:figs}
\end{figure*}

A full simulation of neutral and charged
current inclusive cross section
measurements, including first systematic error considerations,
has been performed assuming one year
of data at maximum luminosity \cite{max:dis07}. 
Integrated luminosities of $1 \ {\rm fb^{-1}}$ at low $Q^2$ and
$10 \ {\rm fb^{-1}}$ at high $Q^2$ are assumed (see section~\ref{geometry}),
leading to 
uncertainties at the $1 - 3 \%$ level over much of the kinematic plane
after a single year of data taking.
The resulting 
cross sections have been used 
as input to an NLO DGLAP \cite{dglap} QCD fit,
similar to those used by the HERA collaborations, in order to
estimate the obtainable LHeC precision on the parton distributions.  

LHeC data could separately constrain 
all of the quark flavours for the first time
in a single experiment. The large luminosities
would give a much improved $x F_3$ measurement compared with HERA,
and the valence densities could hence be extracted precisely. 
With copious charged current data for both $e^+ p$ and $e^- p$ 
collisions, up and down quark distributions
could be separated. With state-of-the-art
secondary vertex detection and a small beam-spot 
($35 \times 15 \ {\rm \mu m}^2$
has been assumed), heavy flavour quarks could be identified,
leading to measurements of the 
charm and beauty structure functions $F_2^{c \bar{c}}$ and
$F_2^{b \bar{b}}$ to a few percent over a wide kinematic 
range \cite{max:dis07,olaf}. 
With high luminosity
the $s$ and $\bar{s}$ densities could be measured to unrivalled precision by
tagging charm quarks in charged current scattering ($Ws \rightarrow c$),
potentially casting new light on a possible $s - \bar{s}$
asymmetry \cite{nutev}.
The enhanced sensitivity to $\partial F_2 / \partial \ln Q^2$
due to the larger lever-arm in $Q^2$ throughout the $x$ range
translates directly into a new level of precision on the gluon density. 
In 
the process of fitting QCD evolution equations to LHeC data, 
the strong coupling constant could be measured to an 
unprecedented experimental precision of a few per mille \cite{kluge}. 

Examples of possible LHeC constraints on parton densities are shown in
figure~\ref{pdf:figs}. 
Very large improvements in uncertainties over present gluon 
density extractions are possible
over several orders of magnitude and notably at the largest $x$,
corresponding to the LHC parton-parton energy frontier.
The ratio $d/u$ as $x \rightarrow 1$ is also much more strongly constrained
than hitherto, which may be important in interpreting high mass
LHC signals.

As in the HERA case, 
the best reconstruction of $x$ and $Q^2$
over the full kinematic plane requires the use of the hadronic
final state 4-vector as well as that of the scattered electron. 
At the newly accessed lowest $x$ values at the LHeC,
the interacting quark is scattered towards the central region
of the laboratory frame and can be well reconstructed.
However, as $x$ grows at fixed $Q^2$, the
hadronic final state becomes increasingly strongly boosted
in the outgoing proton direction and its measurement becomes
increasingly difficult. At the largest $x$ values,
the resolution obtained from the scattered electron 
4-vector alone degrades severely,\footnote{More
specifically, the resolution on the inelasticity $y = Q^2 / sx$ degrades
as $1/y$ as $y \rightarrow 0$ for fixed $Q^2$.} 
such that some control over the very forward-going 
hadrons is highly desirable. The need for good hadron
reconstruction over a wide $x$ range is clearer still
for the case of charged current scattering, where only the
hadron method is available.

\subsection{Low $x$ strong interaction dynamics}
\label{lowx}

In the low $x$ region, 
the usually `asymptotically free' quarks of DIS
meet a high background 
density of partons, and various novel effects are
predicted. 
Parton density predictions over much of the kinematic range relevant
to the LHC rely on the assumption that the DGLAP approximation to
QCD evolution may be used to evolve partons from the relatively low
$Q^2$ domain of HERA and fixed target DIS experiments to the larger
scales of the LHC. DGLAP must, however, become invalid
at some low value of $x$, where $\ln 1/x$ terms in the evolution
become important \cite{bfkl,ccfm}
and where resummation approaches may be required \cite{abf}.  
Although no evidence of deviations from DGLAP evolution has been
obtained from fits to inclusive HERA data, there have been hints 
from hadronic 
final state observables such as forward jets \cite{fjets}
and azimuthal
jet decorrelations \cite{jet:decor},
which are sensitive to the transverse momentum
ordering patterns in the parton cascade.
It is possible that more will be learned from fitting
inclusive LHeC data, though direct observation of the final state parton
cascade will remain necessary for a complete picture, requiring
hadronic final state acceptance at very small angles 
to the beampipe \cite{hannes}.

At sufficiently low $x$, 
unitarity constraints become important and a `black body' limit is
approached \cite{gribov}, in which the cross section 
reaches the geometrical bound given by the transverse proton
size. 
This limit is characterised by $Q^2$ dependences which differ
fundamentally from the usual logarithmic
scaling violations, diffractive cross sections approaching
$50\%$ of the total and other new effects \cite{strikman}.
Applying the black body bound to
the inelastic cross section for the interaction of a 
colour dipole, formed from a $\gamma^* \rightarrow q \bar{q}$ splitting,
leads to an approximate constraint on the gluon 
density $x g(x, Q^2) < Q^2 / \alpha_s$ \cite{gluon:constrain}, 
only a small factor beyond predictions 
for the gluon at the lowest LHeC $x$
values.   

Whilst we have some understanding of the physics of the black body limit 
itself,
the mechanisms and precise dynamics by which it is approached are
completely unknown. `Saturation' effects, in which the low $x$ growth of
the parton densities is tamed must be present at some value
of $x$,
a possible mechanism being 
the onset of recombination processes such as $gg \rightarrow g$ \cite{glr},
leading to non-linear evolution. 

Although no significant saturation signals have been observed
in parton density fits to HERA data, hints  
have been obtained by fitting the data to dipole 
models \cite{jeff,soyez,dipoles}, 
which may be applied at very low $Q^2$
values, beyond the range in which quarks and gluons can be considered
to be good degrees of freedom.
The typical conclusion \cite{jeff} is that
HERA data in the perturbative regime do not exhibit any 
evidence for saturation.
However, when data in the $Q^2 < 1 \ {\rm GeV^2}$
region are included, 
only models which include saturation effects are successful. 
Whether or not this low $Q^2$ HERA saturation effect is confirmed,
it is desirable to fully understand the mechanisms
behind saturation in terms of parton dynamics, 
which may be possible by studying the very low $x$
region at somewhat larger $Q^2$ at the LHeC.

\begin{figure}[h] \unitlength 1mm
  \begin{center}
    \begin{picture}(50,70)
      \put(-16,-10){\epsfig{file=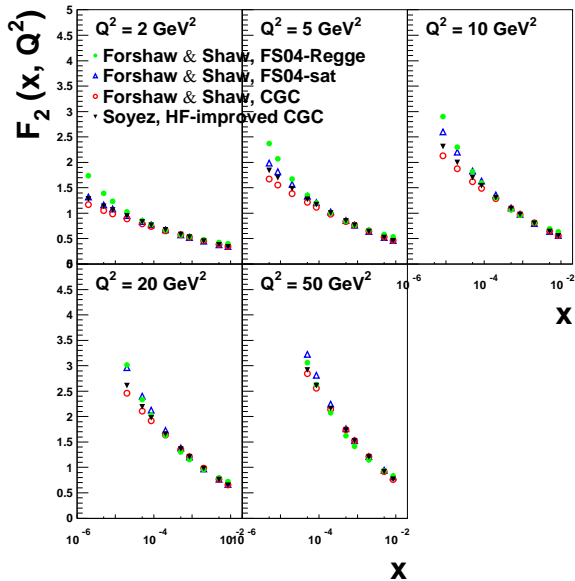,height=0.475\textwidth}}
    \end{picture}
  \end{center}
  \caption[]{Extrapolations into the LHeC regime of different dipole
models \cite{jeff,soyez} 
based on fits to HERA data. The results are shown as pseudo-data
points corresponding to the simulated LHeC measurement \cite{max:dis07}. 
The cross section 
for the dipole-proton interaction
contains saturation effects in all models except for 
`FS04 Regge' \cite{pn:divonne}.}
\label{f2lowx}
\end{figure}

Figure~\ref{f2lowx} shows extrapolations of dipole models constrained by
fits to HERA data to predict the structure function
$F_2(x, Q^2)$ in the LHeC kinematic range. 
The various predictions
are shown as pseudo-data points corresponding to the simulated neutral
current LHeC measurement (section~\ref{partons}).
At the lowest $x$ and $Q^2$, there is a clear distinction between 
the `FS04-Regge' model \cite{jeff}, which does not include saturation effects, 
and all others \cite{jeff,soyez}, 
which include saturation as estimated from low $Q^2$
HERA data. However, any such sensitivity is lost by around
$Q^2 = 50 \ {\rm GeV^2}$, emphasising the importance of low angle 
scattered electron acceptance. 

Whilst such extrapolations give encouraging indications, the 
unequivocal establishment
of parton saturation is more complex. Two studies 
using very different approaches to PDF fitting are in
progress \cite{jeff:fits,NNPDF,pn:divonne}. They both subject
LHeC pseudo-data based on dipole models
to NLO DGLAP fits,
to determine whether saturation effects could be masked,
for example by a sufficiently
flexible parton parameterisation.
It is not yet clear whether 
a breakdown
of pure DGLAP dynamics 
may be visible with $F_2$ data alone. If not, the 
addition of $F_L$ data as a second observable in the fits is likely to
prove conclusive. 




\subsection{Diffraction}
\label{diffraction}

Non-inclusive observables promise to enhance the
LHeC sensitivity to non-linear evolution
and saturation phenomena.
Diffractive channels are promising, since the 
underlying exchange of a pair
of gluons may enhance the sensitivity compared with the single
gluon involved in inclusive processes.
The cleanest processes experimentally
are Deeply-Virtual Compton Scattering 
(DVCS, $ep \rightarrow e \gamma p$)
and vector meson production ($ep \rightarrow e Vp$), 
which have played a major role at HERA \cite{marage}, but
where no detailed LHeC work has been done to date  \cite{pn:dis07}.
 
\begin{figure*}[tb] \unitlength 1mm
  \begin{center}
    \begin{picture}(100,50)
      \put(-25,7){\epsfig{file=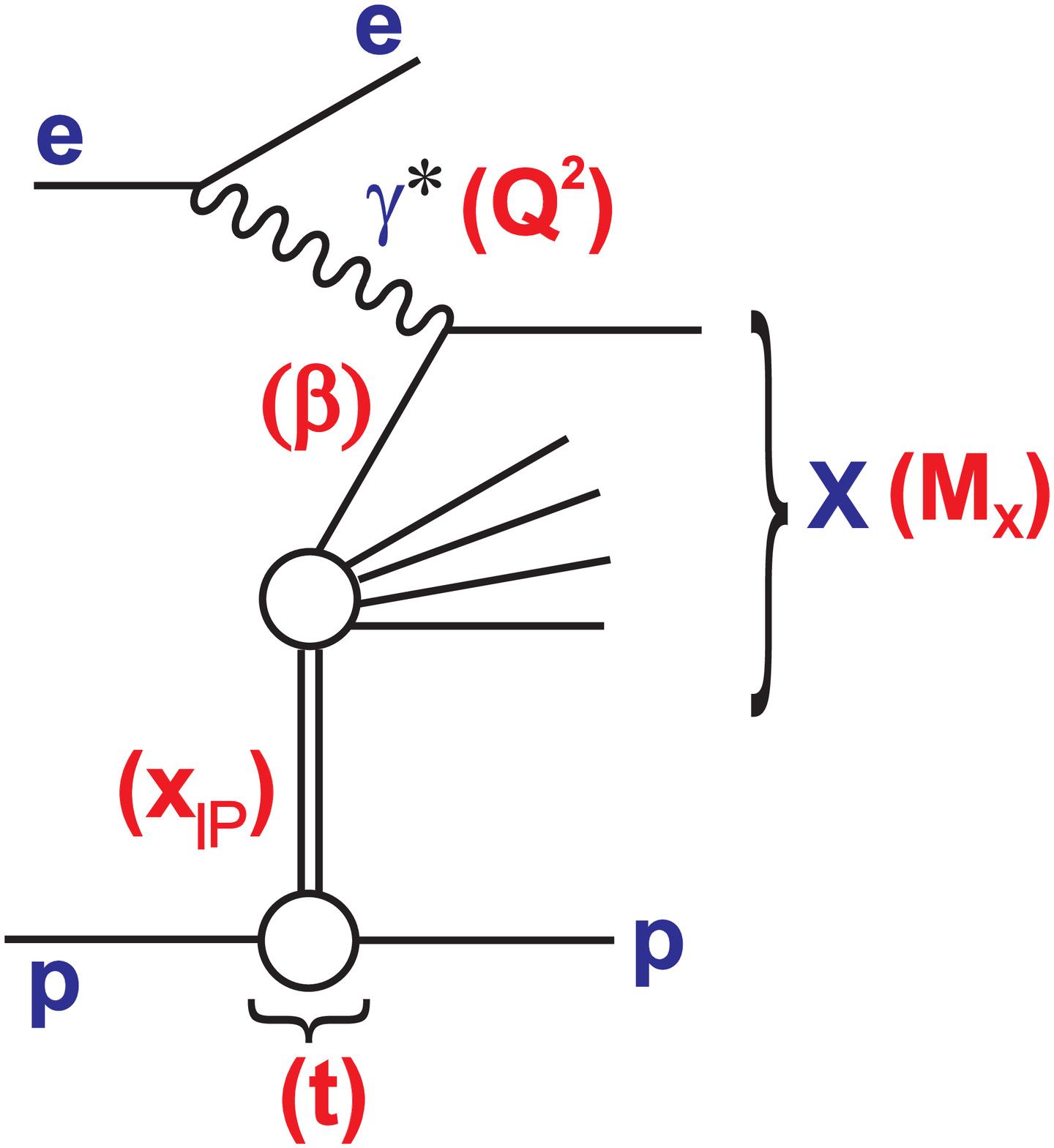,height=0.23\textwidth}}
      \put(4,-8){\epsfig{file=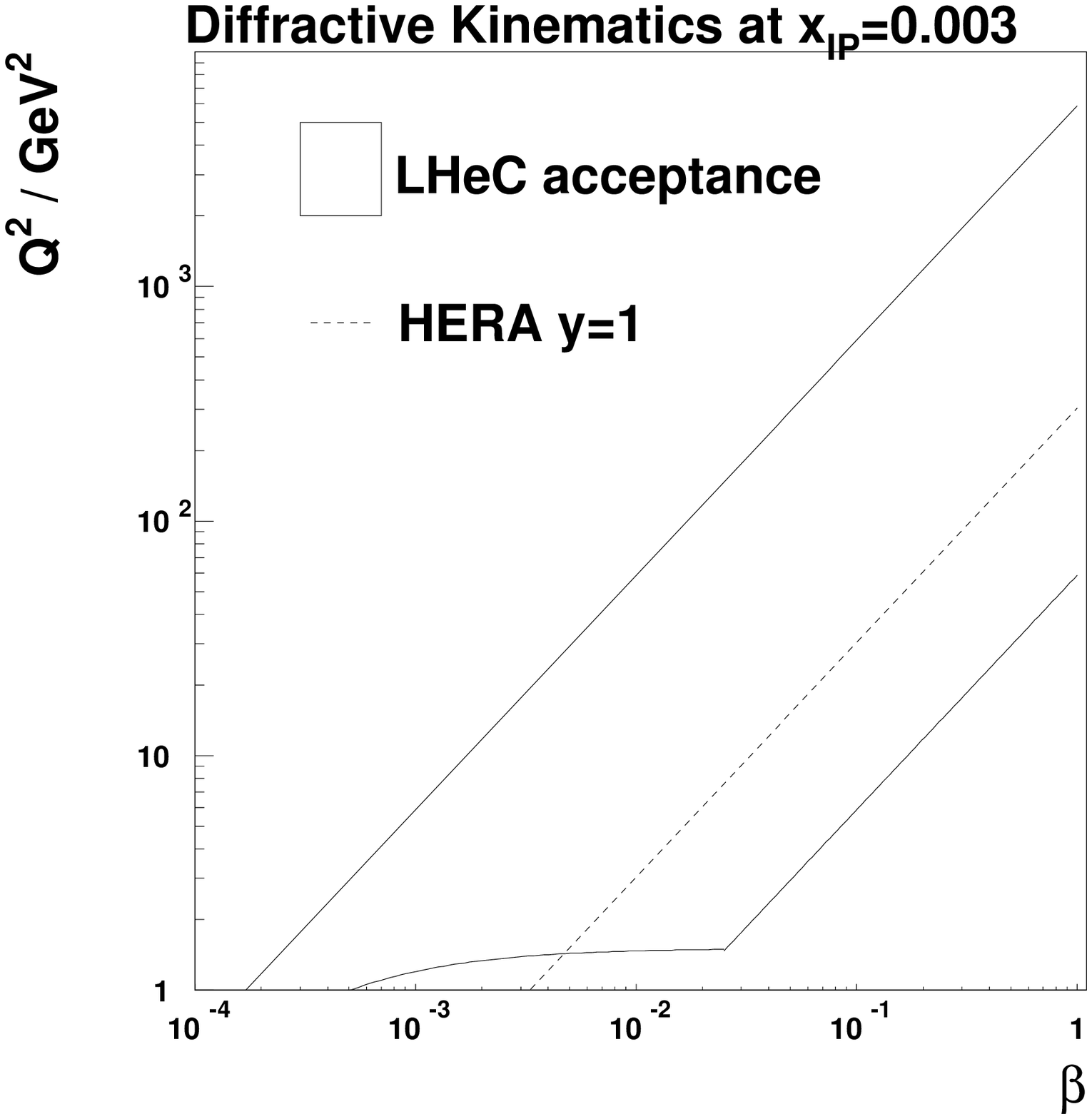,height=0.42\textwidth}}
      \put(65,-8){\epsfig{file=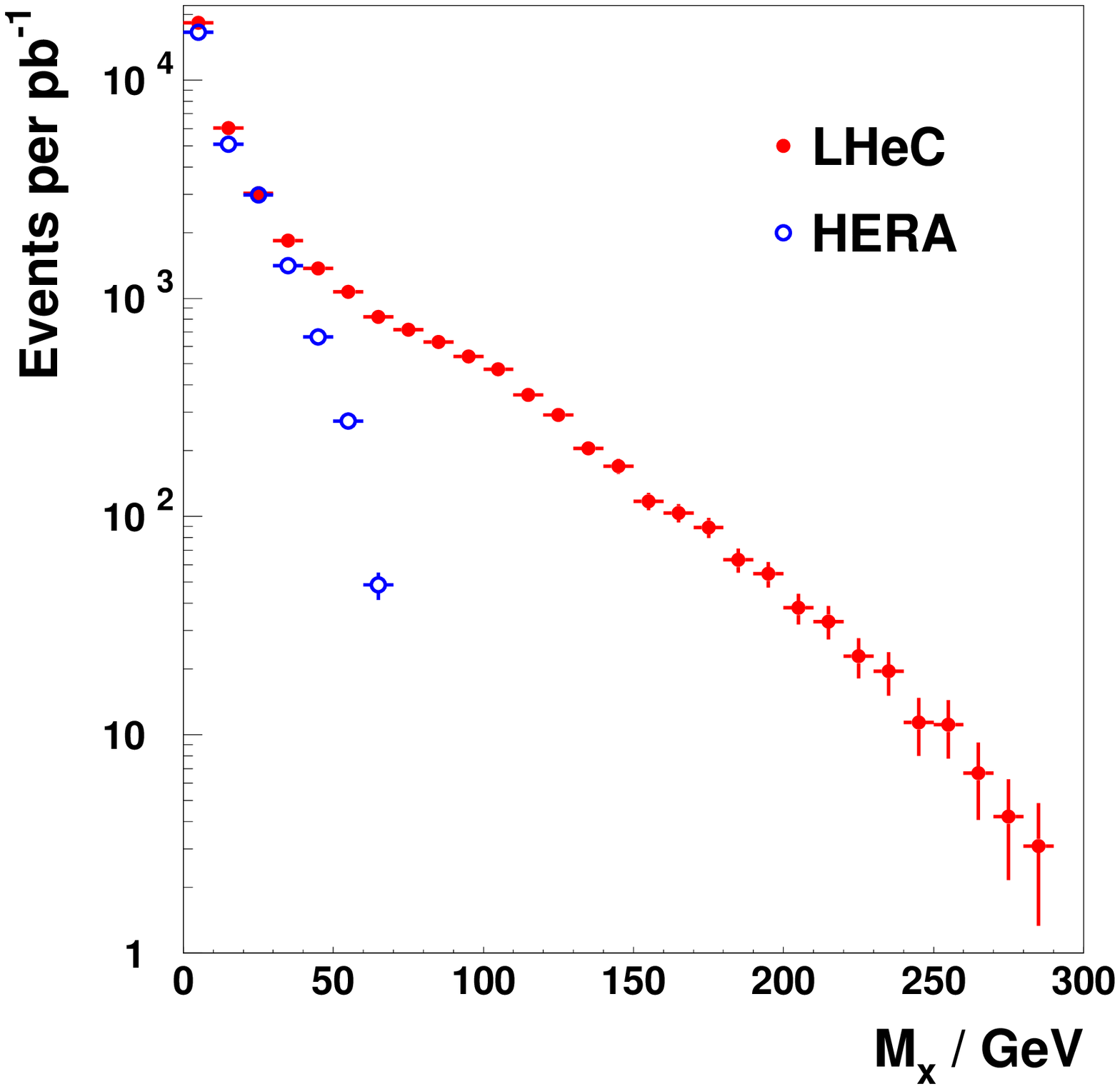,height=0.42\textwidth}}
      \put(-26,45){\Large{\bf{(a)}}}
      \put(22,22){\Large{\bf{(b)}}}
      \put(98,45){\Large{\bf{(c)}}}
    \end{picture}
  \end{center}
  \caption[]{(a) Illustration of the diffractive DIS process, 
$ep \rightarrow eXp$. (b) Comparison between the accessible
kinematic plane in $\beta$ and $Q^2$ at HERA and the LHeC for 
$x_{_{I\!\!P}} < 0.003$. The LHeC phase space is defined
by $0.01 < y < 1$ and $\theta_e < 179^\circ$.
(c) Comaparison between diffractive masses $M_X$ produced at 
HERA and the LHeC with 
$x_{_{I\!\!P}} < 0.05$ \cite{pn:dis07}.}
\label{f2dfigs}
\end{figure*}

First studies \cite{pn:divonne}
have been made of the possibilities with inclusive 
diffraction, $ep \rightarrow eXp$ (figure~\ref{f2dfigs}a),
An impression of the extension in kinematic
coverage at the LHeC 
is given in figure~\ref{f2dfigs}b for an example 
fractional scattered proton energy loss of $x_{_{I\!\!P}} = 0.003$. 
Similarly
to inclusive DIS, fractional struck quark
momenta relative to the diffractive exchange, $\beta = x / x_{_{I\!\!P}}$, 
a factor of around 20 lower than at 
HERA are accessible at the LHeC.

Large improvements in diffractive parton densities (DPDFs) \cite{h1:f2d}
are possible from NLO DGLAP fits to diffractive structure function
data. The extended phase space towards large $Q^2$ at fixed $x$ increases the
lever-arm for extracting the diffractive gluon density and opens the
possibility of significant weak gauge boson exchange, which would allow a
quark flavour decomposition for the first time. Figure~\ref{f2dfigs}c
shows a comparison between HERA and the LHeC in the invariant masses 
$M_X$ produced in diffractive processes with $x_{_{I\!\!P}} < 0.05$ 
(RAPGAP Monte Carlo model \cite{rapgap}). 
Diffractive masses
up to several hundred ${\rm GeV}$ are accessible, such that diffractive final
states involving beauty quarks and $W$ and $Z$ bosons, or even exotic
states with $1^-$ quantum numbers, would be
produced. In addition, diffractive jet and heavy flavour final states 
could be studied at much larger transverse momenta than previously,
reducing the dominant theory scale uncertainties and thus 
providing precision tests of factorisation properties 
and measurements of the problematic gluon density at large $\beta$.

Leading twist diffraction has been related \cite{gribov,diff:shadow} to the 
leading twist component 
of the nuclear shadowing phenomenon, in which the exchanged virtual
photon in an $eA$ interaction scatters coherently from more than one
nucleon. Measuring diffractive DIS together with nuclear structure
functions (section~\ref{ions}) in the LHeC range therefore tests the
unified picture of complex strong interactions and leads to a 
detailed understanding of the shadowing mechanism, possibly 
essential in interpreting saturation signatures in $eA$ interactions.

The strong forward boost of the hadronic final state for most processes
at the LHeC will limit the use of the standard `large rapidity gap'
technique for selecting diffractive events 
to lower $x_{_{I\!\!P}}$ values than in the HERA 
case.\footnote{Expressed in terms of the most forward
extent $\eta_{\rm max}$ 
of the $X$ system,
$x_{_{I\!\!P}} = 0.05$ corresponds to
$\eta_{\rm max} \sim 3.5$ at HERA and
$\eta_{\rm max} \sim 5.5$ at the LHeC.} 
Studying inclusive diffraction at the 
LHeC may therefore best be done by directly detecting scattered protons. 
Given that the beam optics for outgoing protons
at the LHeC would not be very different from those at the LHC,
first considerations \cite{pvm} suggest that that
the region around $420 \ {\rm m}$ from the interaction point
would be a suitable place to install a beamline proton spectrometer,
similar to that under consideration by ATLAS and CMS \cite{fp420}.
First studies
have also been made of leading neutron cross sections 
at the LHeC \cite{pn:divonne},
where designs similar to the `leading neutron' or `zero degree'
calorimeters at HERA and LHC experiments may be appropriate. 

\subsection{Heavy ion physics}
\label{ions}

Whilst establishing parton saturation in $ep$ collisions at the LHeC 
may require multiple observables, 
more striking signals may be available in
$eA$ interactions. The small $x$ nuclear
gluon density $g_A$ at central impact parameters
is enhanced relative to that ($g_N$) in a nucleon by a factor
$(g_A / \pi R_A^2) / (g_N / \pi R_N^2) \simeq A^{1/3} g_A / A g_N
\simeq A^{1/3}$ \cite{strikman}, where 
$R_N$ and $R_A$ represent the nucleon and nuclear radii, respectively, and
nuclear shadowing is neglected.  
This corresponds to a factor of around 6 for the lead ions which will
be used at the LHC, leading to gluon densities close to estimates
of the black body limit (section~\ref{lowx}).
If a really clear signature for parton saturation exists,
for example a dramatic flattening of the $x$ dependence of the 
$F_2$ structure function at low $x$ or anomalously strong 
scaling violations, 
it is likely to be established
first in $eA$ collisions, or through comparisons of $eA$ with $ep$ data.

\begin{figure}[h] \unitlength 1mm
  \begin{center}
    \begin{picture}(50,57)
      \put(-12,-10){\epsfig{file=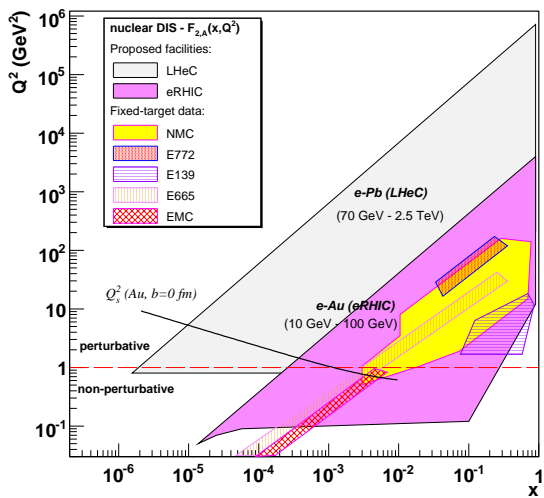,height=0.425\textwidth}}
    \end{picture}
  \end{center}
  \caption[]{Kinematic plane showing previous \cite{nmc} and possible 
future (EIC \cite{eic} and LHeC) coverage for $eA$ collisions \cite{dde}.}
\label{eA:kinematics}
\end{figure}

Experimentally, scattering leptons from the LHC heavy ion beams 
at large $Q^2$ is
expected to lead to final states which do not look dramatically different
from $ep$ scattering. However, unlike for the $ep$ case, colliding beam
configurations have never previously been employed for $eA$ collisions.
Our current knowledge of nuclear parton 
distributions is thus restricted to the phase space covered by fixed target
experiments as shown in 
figure~\ref{eA:kinematics}. For example, the NMC data 
in the DIS region barely extend
below $x = 10^{-2}$ and are restricted to relatively light nuclei 
(helium, carbon and calcium) \cite{nmc}. The LHeC would 
extend the known range by up to four orders of magnitude, 
offering for the first time a quantitative understanding of the 
(presumably saturated)
initial state partons entering the heavy ion collisions which 
are expected to produce quark gluon plasma conditions at the LHC. 

Figure~\ref{eA:kinematics} also shows the extrapolation of the 
critical saturation line at central impact parameter, 
estimated from dipole model fits to 
low $Q^2$ HERA data (section~\ref{lowx}). With sufficient
acceptance at low polar angles, the LHeC data falls well beyond this line,
whilst remaining in the perturbative region.  

In addition to the programme with lead or other heavy ions, 
neutron structure could be explored at the LHeC through electron-deuteron 
scattering \cite{hera3}, 
as has been done in many fixed target DIS experiments.
This adds complementary information in PDF extractions assuming
isospin conservation
and thus allows QCD evolution studies of valence parton 
densities. The Gott\-fried sum rule \cite{gottfried} could also
be studied, measuring 
the difference $\bar{u} - \bar{d}$ and
hence testing previous observations \cite{nmc:obs} of
a light flavour asymmetry in the nucleon sea. 

\section{MACHINE AND DETECTOR}
\label{experiments}

The challenge to realise the 
wide-ranging physics possibilities of the LHeC
is to collide the LHC protons or heavy 
ions with a new electron beam at high luminosity, without 
inhibiting the ongoing LHC hadron-hadron collision programme.
The most promising locations for the interaction point are 
on the sites of the current ALICE and LHCb experiments, after they
have completed their programmes. 
The principle
limitation on the achievable electron energies and luminosities is power
consumption, which is large for electron accelerators due to 
synchrotron radiation losses. 
A working limit of $100 \ {\rm MW}$ wall-plug power
is assumed. 


Two basic lay-outs are being considered for the electron 
accelerator \cite{bruening}. 
An electron beampipe in the same tunnel as the LHC has the advantage 
of high expected luminosity \cite{ferdi,burkhardt:dis08,epac08:rr}, 
though the achievable electron 
beam energy is limited by the synchrotron power, which grows as
$E_e^4$. For acceptable power consumption, a luminosity of around
$3 \times 10^{33} {\rm cm^{-2} s^{-1}}$ might be achievable at 
$E_e = 50 \ {\rm GeV}$, corresponding to an integrated
luminosity of $20 \ {\rm fb^-1}$ per year. A $70 \ {\rm GeV}$
electron beam would yield a factor of around four less. 
Synchronous $ep$ and $pp$ LHC operation appears to be possible, with
by-pass tunnels around existing experiments of length 
a few hundred metres \cite{burkhardt}
also housing around 100 cavities and klystrons 
comprising the electron beam RF infrastructure \cite{linnecar}. 
The tunnels could be excavated in parallel 
with LHC $pp$ operation. A 
finite crossing angle is required
in order to ensure that the outgoing electron beam
does not `parasitically' 
interfere with the proton bunch arriving for the next bunch 
crossing. It may be possible to avoid 
significant resultant luminosity losses 
by using crab cavities \cite{crabs}.

Injection to an electron ring\footnote{The LEP injector complex and
the space used for its RF components
are no longer available.} could be 
provided by the Superconducting Proton Linac (SPL), which is under 
consideration as part of the LHC injection chain upgrade \cite{garoby}. 
An initial phase of the 
LHeC could even use multiple passes in the 
SPL for the full electron acceleration, 
producing beam energies of around $20 \ {\rm GeV}$.

An alternative electron beam solution is a linear 
accelerator (linac), arriving
tangentially at an LHC interaction point. 
The linac could use ILC cavity technology 
in pulsed or continuous wave mode. 
This approach has the advantage
of construction being relatively 
decoupled from the LHC proton ring and
energies of $E_e = 100 \ {\rm GeV}$ and beyond have been 
discussed \cite{bruening}, corresponding to electron-quark collisions 
at a centre of mass energy approaching $2 \ {\rm TeV}$.
The final focus could be further from the interaction point than for the
case of an electron ring, improving low angle detector acceptance. 
In scenarios sketched to 
date \cite{epac08:lr,zimmermann}
which have acceptable power consumption, the linac option leads to 
lower luminosity than for an electron ring, which has led to proposals
to exploit energy recovery techniques.
Since a linac-ring configuration would represent a completely new approach
to high energy colliders, a large amount of research and development would
be necessary.


Detailed calculations of the 
LHeC electron beam optics have led to proposals for the lay-out of
the interaction region \cite{holzer}, which is also 
a major consideration for the detector design. 
The highest 
projected luminosities
are achieved by placing beam 
focusing magnets close to the interaction point. 
On the other hand, 
there are a number of reasons
(section~\ref{physics}) why hermetic instrumentation for electron and
hadronic final state detection is a highly desirable feature of a DIS
detector. 
A number
of novel solutions to the compromise between acceptance and luminosity
have been proposed, including a 2 stage approach, similar to
HERA-I and HERA-II, and the possibility of having two interaction points,
focused respectively on high $Q^2$ and 
low $x$ physics. More exotic possibilities
involve integrating the electron beam focusing and deflecting
magnets into the detector design. An example is to
instrument the superconducting
focusing magnets to give a calorimetric 
response \cite{greenshaw} by exploiting scintillation light
in the liquid helium 
of the cryogenics
produced by charged particle components of showers.


A first detector overview for $ep$ and $eA$ physics 
at the LHeC has been sketched. Although it is too early to 
decide on technologies, promising initial lay-outs 
have been suggested \cite{kostka}, 
which feature high resolution on the scattered electron and
the hadronic final state and a relatively low material budget,
leading towards a new level of precision in DIS.
Central tracking and vertexing devices 
might be based on various pixel technologies, whilst 
CALICE-type solutions and liquid argon sampling are under
consideration for electromagnetic and hadronic calorimetry, respectively.
A possible approach to the magnetic fields \cite{kostka:private}
is a double solenoid, which could yield
$\sim 5 \ {\rm T}$ in the tracking region, $\sim 1.5 \ {\rm T}$
in the muon area and would not require a return yoke.

\begin{figure*}[tb] \unitlength 1mm
  \begin{center}
    \begin{picture}(100,110)
      \put(-5,-10){\epsfig{file=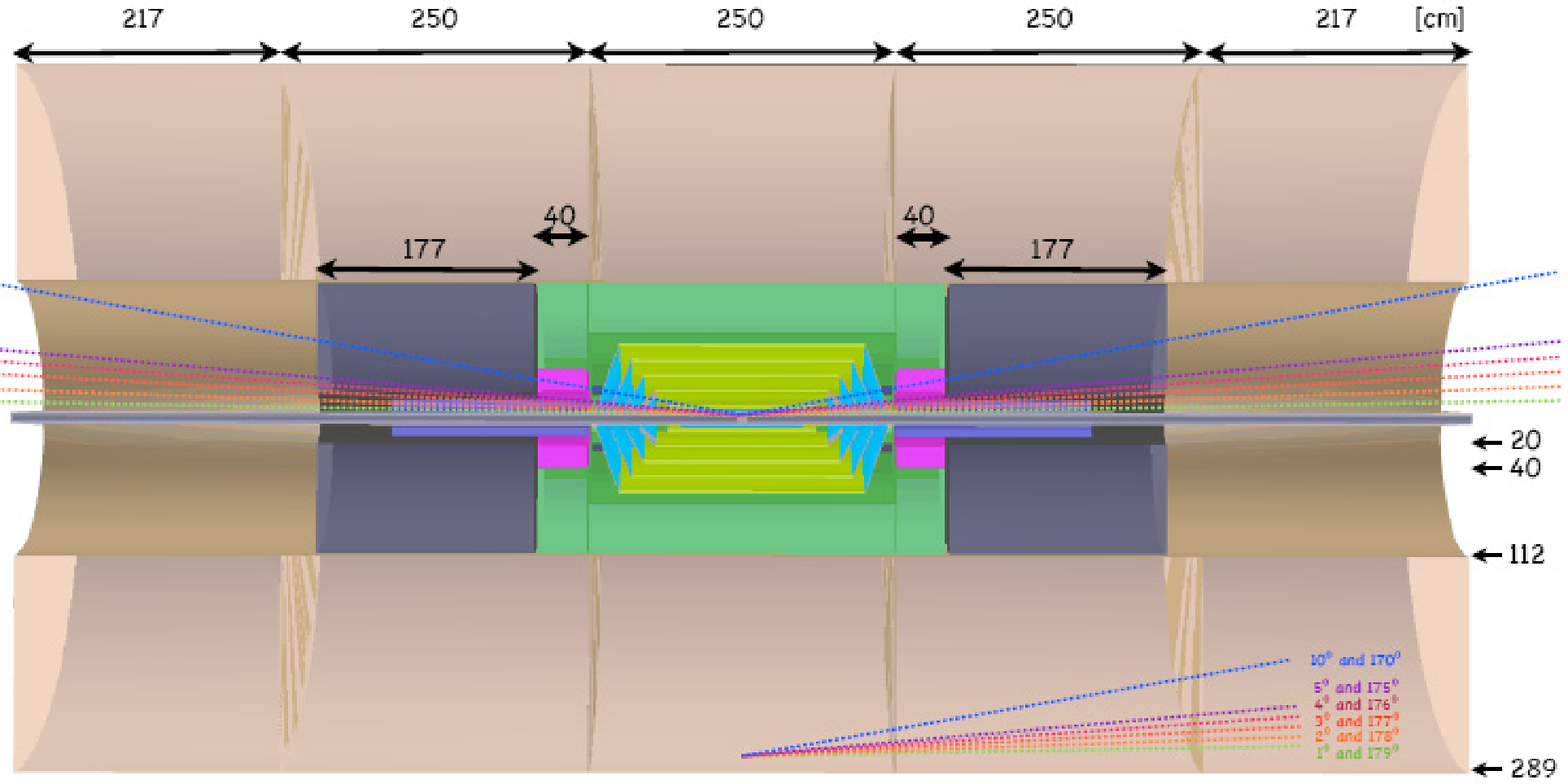,height=0.38\textwidth}}
      \put(-5,53){\epsfig{file=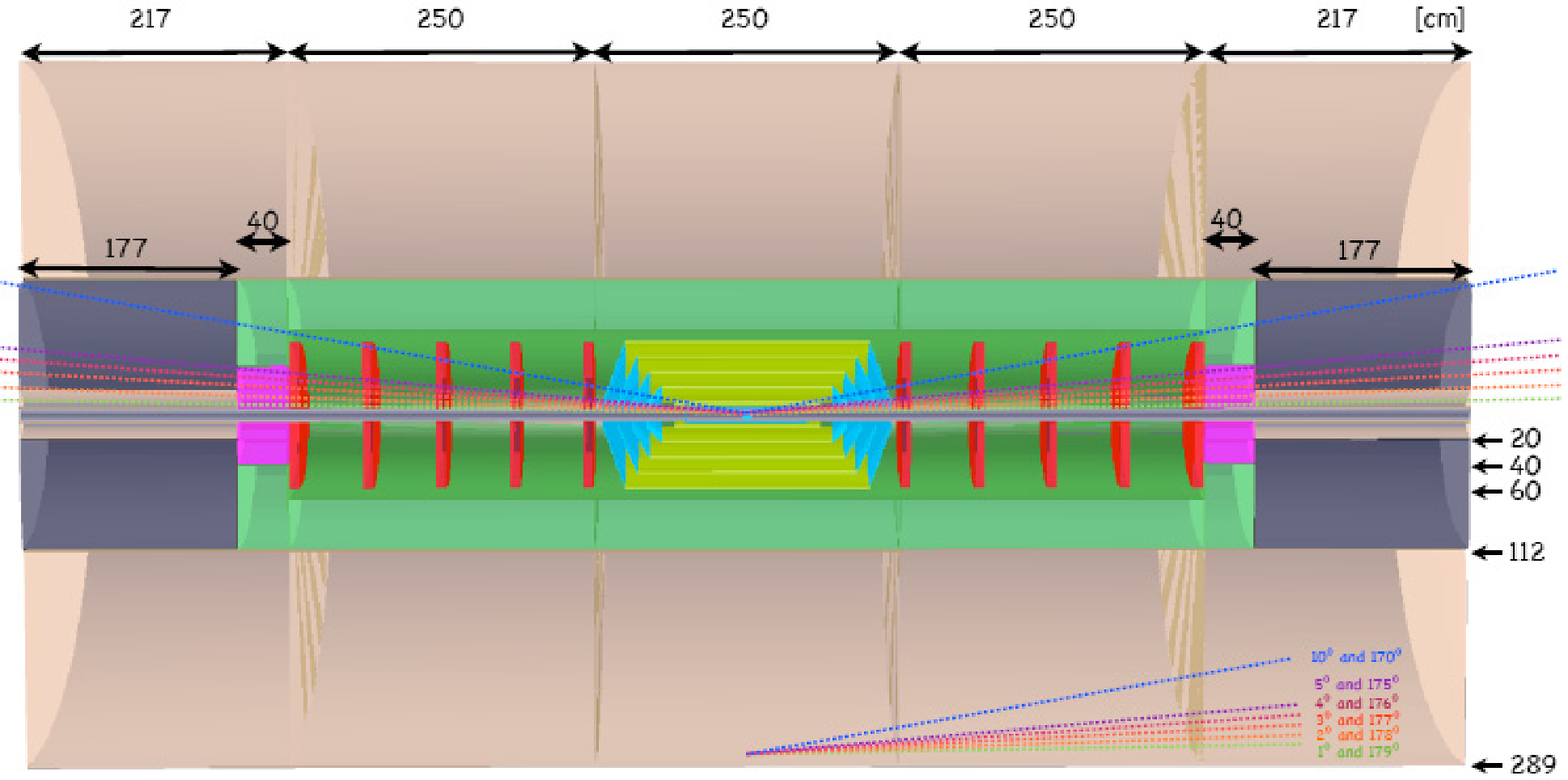,height=0.38\textwidth}}
      \put(-26,80){\Large{\bf{(a)}}}
      \put(-26,18){\Large{\bf{(b)}}}
    \end{picture}
  \end{center}
  \caption[]{Sketches of detector lay-outs 
with tracking acceptance covering 
(a) $1 < \theta < 179^\circ$ and (b) 
$10 < \theta < 170^\circ$ \cite{kostka}. In the modular design shown, 
the extended tracker in (a) may be removed to provide space for the
final beam focusing elements, with calorimeter inserts also moving
closer to the interaction point (b). Dimensions are given in {\rm cm}
and angles from $1^\circ$ to $10^\circ$ are indicated.} 
\label{detector:fig}
\end{figure*}

With a beam-pipe of cross sectional dimensions a few ${\rm cm}$,
giving sufficient space for the synchrotron radiation fan, a
particle scattered at a polar angle of $1^\circ$ to the beam does not
emerge from the beampipe until it has travelled a
longitudinal distance of
a few metres. Measuring such tracks
therefore requires a long tracking detector, as illustrated for 
example in figure~\ref{detector:fig}a. In the example
modular design depicted, there are movable 
calorimeter inserts,
which sit behind the tracking region in the low $x$ configuration
and move closer to the interaction region when beam focusing magnets
and a shorter tracking region are introduced (figure~\ref{detector:fig}b).
Other suggested solutions to particle detection at low scattering angles
are instrumenting 
inside the electron beam-pipe and introducing a dipole field beyond the 
interaction region to sweep out particles scattered
through small angles \cite{caldwell}.

\section{SUMMARY}

A new
investigations of the possibility of exploiting 
the LHC proton beams for $ep$
physics is well underway in the framework of the LHeC workshop.
First, promising, though often crude, evaluations of the physics 
potential are in place and
detector, interaction region and 
accelerator lay-out possibilities are being debated. 
Solutions with an electron ring or linac and with or without
final focusing beam elements near to the interaction point are 
all being pursued, in order to 
understand fully the advantages and consequences of each. 

Following an interim report presented to ECFA in
November 2008 \cite{max:ecfa}, work towards a
Conceptual Design Report is ongoing. Frequent updates can be
found at \cite{LHeC:web}. 
The aim is to produce the report by early 2010, to be used as input to
pursuant CERN strategy discussions.
If realised, the LHeC facility would become an 
integral part of the quest to fully understand the new Terascale 
physics which will emerge as the LHC era unfolds.

\section{ACKNOWLEDGEMENTS}

Thanks to all colleagues who have contributed to the LHeC project
so far, to the organisers of the Ringberg meeting for 
giving me the opportunity of summarising and presenting this work
and to Oliver Br\"{u}ning, G\"{u}nther Grindhammer, Max Klein
and Peter Kostka for providing valuable comments to 
a draft version of this document.


\newpage

\begin{thebibliography}{99}
\bibitem{reviews} For recent summaries, see H. Abramowicz, 
{\em `Selection of ZEUS results'}, doi:10.3360/dis.2008.2,
and A. Sch\"{o}ning,
{`New Results from the H1 Collaboration'}, doi:10.3360/dis.2008.1,
Proc. of DIS08 (London);
C. Diaconu, {\em `Structure functions'} and C. Glasman,
{\em `Precision tests of QCD with jets and vector bosons at HERA and 
Tevatron'}, Proc. of ICHEP08 (Philadelphia).   
\bibitem{HERA:LHC} R. Thorne, these proceedings; 
Proc. of the 2nd, 3rd and 4th HERA-LHC workshops, in litt; 
\verb+http://www.desy.de/~heralhc+.
\bibitem{ferdi} J. Dainton et al., JINST {\bf 1}, P10001 (2006).
\bibitem{LHeC:web} The LHeC project web page: \verb+http://www.lhec.org.uk+.
\bibitem{previous:ep} J. Feltesse, R. R\"{u}ckl and A. Verdier, presentations
at LHC workshop, Aachen, 1990, CERN 90-10 (1990); 
{\em `The THERA Book: $ep$ Scattering at $\sqrt{s} \sim 1 \ {\rm TeV}$'},
(DESY 01-123F vol 4); 
{\em `QCD Explorer Based on LHC and CLIC-I'}, D. Schulte
and F. Zimmermann, presentation at EPAC'04 (Lucerne), CERN-AB-2004-079 (2004).
\bibitem{eic} The Electron Ion Collider Web Page:
\verb+http://web.mit.edu/eicc/+.
\bibitem{divonne} 1st ECFA-CERN LHeC Workshop web page:
\verb+http://indico.cern.ch/+ \\
\verb+conferenceDisplay.py?confId=31463+.
\bibitem{max:dis07} M. Klein, {\em `Parton Distributions at the LHeC'},
doi:10.3360/dis.2007.215, Proc. of DIS07, Munich (2007).
\bibitem{salam} A. Salam, {\em `The Unconfined Quarks and Gluons'},
Proc. of 18th Rochester Conference, Tbilisi (1976).
\bibitem{dainton} J. Dainton, {\em `The Large Hadron-Electron Collider'},
doi:10.3360/dis.2007.225, Proc. of DIS07, Munich (2007).
\bibitem{zarnecki} A. Zarnecki, {\em Leptoquarks and Contact 
Interactions at LHeC'}, 
doi:10.3360/dis.2008.234, Proc. of DIS08, London (2008).
\bibitem{buchmueller} W. Buchm\"{u}ller, R. R\"{u}ckl and D. Wyler,
Phys. Lett. {\bf B191}, 442 (1987); erratum ibid {\bf B448}, 320 (1999).
\bibitem{trinh} N. Trinh {\em `Excited Fermions}, see \cite{divonne}.
\bibitem{selectron} E. Perez, {\em `Physics Beyond the 
Standard Model at the LHeC'}, 
doi:10.3360/dis.2007.219, Proc. of DIS07, Munich (2007).
\bibitem{top} G. Brandt, {\em `Single Top Production'},
see \cite{divonne}.  
\bibitem{uta} U. Klein, {\em `Higgs Cross Sections at the LHeC'}
and M. Khoze {\em `Backgrounds to Higgs Production at the LHeC'},
see \cite{divonne}. 
\bibitem{dglap}V. Gribov and L. Lipatov, Sov. J. Nucl. Phys. {\bf 15},
438 \& 675 (1972); 
L. Lipatov, Sov. J. Nucl. Phys. {\bf 20}, 94 (1975); 
G. Altarelli and G. Parisi, Nucl. Phys. {\bf B126},
298 (1977); 
Y. Dokshitzer, Sov. Phys. JETP {\bf 46}, 641 (1977).
\bibitem{cteq6} D. Stump et al., JHEP {\bf 0310}, 046 (2003).
\bibitem{olaf} O. Behnke, {\em `Precision Investigations of QCD
and Electroweak Interactions'}, see \cite{divonne}.  
\bibitem{nutev} NuTeV Collaboration, Phys. Rev.  {\bf D64}, 112006 (2001).
\bibitem{kluge} T. Kluge, {\em `Prospects for $\alpha_s$ determination 
in DIS'}, doi:10.3360/dis.2008.233, Proc. of DIS08, London (2008).
\bibitem{bfkl} V. Fadin, E. Kuraev and L. Lipatov, Sov. Phys. JETP
{\bf 44}, 443 (1976); 
V. Fadin, E. Kuraev and L. Lipatov, Sov. Phys. JETP
{\bf 45}, 199 (1977); 
Y. Balistsky and L. Lipatov, Sov. J. Nucl. Phys.
{\bf 28}, 822 (1978).
\bibitem{ccfm} M. Ciafaloni, Nucl. Phys. {\bf B296}, 49 (1988); 
S. Catani, F. Fioriani and M. Marchesini,
Phys. Lett. {\bf B234}, 339 (1990); 
S. Catani, F. Fioriani and M. Marchesini,
Nucl. Phys. {\bf B336}, 18 (1990); 
M. Marchesini, Nucl. Phys. {\bf B445}, 49 (1995).
\bibitem{abf} G. Altarelli, R. Ball and S. Forte, 
Nucl. Phys. {\bf B799}, 199 (2008).
\bibitem{fjets} H1 Collab., Phys. Lett. {\bf B542}, 193 (2002).
\bibitem{jet:decor} H1 Collab., Eur. Phys. J. {\bf C33}, 477 (2004).
\bibitem{hannes} H. Jung, {\em `Small $x$ Parton Dynamics},
see \cite{divonne}. 
\bibitem{gribov} V. Gribov, Sov.\ Phys.\ JETP {\bf 30}, 709 (1970).
\bibitem{strikman} L. Frankfurt et al., 
{\em `Electron-Nucleus Collisions at THERA'}, in THERA 
Book (see \cite{previous:ep}), [hep-ph/0104252].
\bibitem{gluon:constrain} L. Frankfurt, W. Koepf and M. Strikman, 
Phys. Rev. {\bf D54}, 3194 (1996); 
E. Gotsman et al., J. Phys. {\bf G27}, 2297 (2001).
\bibitem{glr} V. Gribov, E. Levin, G. Ryskin, Phys. Rept. {\bf 100}, 1 (1983).
\bibitem{jeff} J. Forshaw and G. Shaw, JHEP {\bf 0412}, 052 (2004)
\bibitem{soyez} G. Soyez, Phys. Lett. {\bf B655}, 32 (2007).
\bibitem{dipoles} K. Golec-Biernat, M. W\"{u}sthoff, 
Phys. Rev. {\bf D59}, 014017 (1999);
E. Iancu, K. Itakura and S. Munier, Phys. Lett. {\bf B590}, 199 (2004);
H. Kowalski, L. Motyka and G. Watt, Phys. Rev. {\bf D74}, 074016 (2006).
\bibitem{jeff:fits} J. Forshaw, {\em `Saturation at the LHeC'},
doi:10.3360/dis.2008.235, Proc. of DIS08, London (2008). 
\bibitem{NNPDF} J. Rojo-Chacon, {`Lox $x$ Physics at LHeC with NNPDFs'},
see \cite{divonne}.
\bibitem{pn:divonne} P Newman, {\em `Physics at High Parton Densities'},
see \cite{divonne}.
\bibitem{marage} A. Bruni, X. Janssen and P. Marage, {\em `Exclusive
Vector Meson Production and Deeply Virtual Compton Scattering at HERA'},
see \cite{HERA:LHC}.
\bibitem{pn:dis07} P. Newman, {\em `Low $x$ Physics at the LHeC'},
doi:10.3360/dis.2007.223, Proc. of DIS07, Munich (2007).
\bibitem{h1:f2d} H1 Coll., Eur. Phys. J. {\bf C48}, 715 (2006).
\bibitem{rapgap} H. Jung, Comput. Phys. Commun. {\bf 86}, 147 (1995).
\bibitem{diff:shadow} L. Frankfurt and M. Strikman, Eur. Phys. J {\bf A5},
293 (1999).
\bibitem{pvm} P. van Mechelen, private communication.
\bibitem{fp420} FP420 R\&D Coll, hep-ex/0806.0302.
\bibitem{nmc} NMC Coll., Nucl. Phys. {\bf B441}, 3 (1995); 
G. Piller and W. Weise, Phys. Rept. {\bf 330}, 1 (2000). 
\bibitem{dde} D. d'Enterria, hep-ex/0706.4182.
\bibitem{hera3} T. Alexopoulos et al., {\em `Electron-deuteron Scattering
with HERA: a Letter of Intent for an Experimental Programme with the
H1 Detector'}, DESY 03-194 (2003).
\bibitem{gottfried} K Gottfried, Phys. Rev. Lett {\bf 18}, 1174 (1967).
\bibitem{nmc:obs} NMC Coll., Phys. Rev. {\rm D50}, 1 (1994).
\bibitem{bruening} O. Br\"{u}ning, {\em `Summary of Accelerator Working 
Group'}, see \cite{divonne}.
\bibitem{burkhardt:dis08} H. Burkhardt, {\em `LHeC Ring-Ring Option'},
doi:10.3360/dis.2008.231, Proc. of DIS08, London (2008).
\bibitem{epac08:rr} F. Willeke et al., {\em `A Storage Ring Based Option
for the LHeC'}, Proc of EPAC08, Genoa (2008).
\bibitem{burkhardt} H. Burkhardt, {\em `Ring-ring layout and by-pass
design'}, see \cite{divonne}.
\bibitem{linnecar} T. Linnecar, {\em `Fitting Electron RF and Power
Sources into the LHC'}, see \cite{divonne}.
\bibitem{crabs} R. Calaga, {\em `Crab Cavities for the LHeC'},
see \cite{divonne}.
\bibitem{garoby} R. Garoby, {\em `LHC injector upgrade plan'},
BEAM'07, Geneva (2007).
\bibitem{epac08:lr} F. Zimmermann et al., {\em `Linac-LHC $ep$
Collider Options'}, Proc of EPAC08, Genoa (2008).
\bibitem{zimmermann} F. Zimmermann, {\em `LHeC Ring-Linac Options'},
see \cite{divonne}.
\bibitem{holzer} B. Holzer, {\em `Interaction Region of the LHeC'},
see \cite{divonne}.
\bibitem{greenshaw} T. Greenshaw, {\em `Active Magnets'},
see \cite{divonne}.
\bibitem{kostka} A. Polini, P. Kostka, R. Wallny, {\em `Detector Working
Group Summary'}, see \cite{divonne}.
\bibitem{kostka:private} P. Kostka, private communication.
\bibitem{caldwell} H. Abramowicz et al., 
{\em `A New Experiment for the HERA Collider: Expression
of Interest'}, MPI-PhE/2003-06 (2003).
\bibitem{max:ecfa} M. Klein, {\em `LHeC: A Large Hadron electron Collider
at CERN'}, report to plenary ECFA, CERN, November 2008.
\end{thebibliography}
\end{document}